\documentstyle[11pt,epsfig,here]{article}
\def\m0{\hbox{$m_{\hbox{\scriptsize 0}}$}}

\def\A0{\hbox{$A_{\hbox{\scriptsize 0}}$}}

\begin{document}
\vspace*{-2cm}
\begin{flushright}
{\bf LAL/RT 00-02}\\
February 2000
\end{flushright}
\vskip 2.5 cm

\begin{center}
{\bf  ANALYTICAL ESTIMATION OF THE DYNAMIC\\
APERTURES OF CIRCULAR ACCELERATORS}
\end{center}
\vspace {1.2 cm}

\begin{center}
{\large \bf Jie Gao}   
\end{center}   

\begin{center}
{\Large\bf Laboratoire de l'Acc\'el\'erateur Lin\'eaire\\}
 {IN2P3-CNRS et Universit\'e de Paris-Sud, BP 34, F-91898 Orsay Cedex}
\end{center}     

\vspace {1 cm}
\begin{abstract}
By considering delta function sextupole, octupole, and decapole
perturbations and using difference action-angle variable equations, 
we find some useful analytical formulae for the estimation of the dynamic apertures of 
circular accelerators due to single sextupole, single octupole, 
single decapole (single 2$m$ pole in general).  
Their combined effects are derived based on the Chirikov criterion
of the onset of stochastic motions. Comparisons with numerical simulations 
are made, and the agreement is quite satisfactory. 
These formulae have been applied to determine the beam-beam limited dynamic 
aperture in a circular collider. 
\end{abstract}

\section{Introduction}
One of the preoccupations of the circular accelerator
designer is to estimate the influence of
nonlinear forces on the single particle's 
motion. These nonlinear forces manifest themselves 
as the systematic and random errors from optical elements, 
the voluntarily introduced functional ones,
such as the sextupoles used for the chromaticity corrections and the octupoles
used for stabilizing the particles' collective motion,
or from nonlinear beam-beam interaction forces. Even though the nonlinear
forces mentioned above compared with the linear forces are 
usually very small, what is 
observed in reality, however, 
is that when the amplitudes of the transverse oscillation of a particle
are large enough, the transverse motion 
might become unstable and the particle
itself will finally be lost on the vacuum chamber. 
Apparently, the above implied maximum oscillation
amplitudes, $A_{x,y}$, corresponding to the stable motions
are functions of the specific longitudinal position, 
$s$, along the machine, and these functions $A_{x,y}(s)$ are 
the so-called dynamic apertures 
of the machine. A reasonably designed machine should satisfy
the condition $A_{x,y}(s)\geq M_{x,y}(s)$, where $M_{x,y}(s)$ are
the mechanical cross-section dimensions of the
vacuum chamber. 
\par
Needless to say the dynamic aperture problem in circular accelerators is one of the
most challenging research topics for accelerator physicists, and the
relevant methods adopted to treat this problem are quite different
analytically and numerically.
In this paper we will show how single sextupole,
single octupole, and single decapole (single 2$m$ pole in general) 
in the machine\linebreak
\newpage
\noindent
 limit the dynamic aperture and what is their combined effect if 
there are more than one nonlinear element.
From the  
established analytical formulae for the dynamic aperture 
one gets the scaling laws
which relate the nonlinear perturbation strengths, beta functions,
and the dynamic apertures.
To test the validity of these formulae,
comparisons with some numerical simulation have been made.
As an interesting application the beam-beam limited dynamic apertures
in a circular collider have been discussed.   
\par
\section{Hamiltonian Formalism}
The Hamiltonian formulation of dynamics is best known not only
for its deep physical and philosophical inspiration to the physicists, but
also for its technical convenience in solving 
various nonlinear dynamical problems. The general Hamiltonian for a particle 
of rest mass, $m_0$, and charge, $e$, in a magnetic vector potential, 
$\bf A$, and electric potential, $\Phi$, is expressed as:   
\begin{equation}
{\cal H}(q,p,t)=e\Phi +c\left(({\bf p}-e{\bf A})^2+m_0c^2\right)^{1/2}
\label{eq:a}
\end{equation}
where $c$ is the velocity of light and $\bf p$ is the momentum
with its components, $p_i$, conjugate to the space coordinates, $q_i$.
The equations of motion can be readily written in terms of Hamilton's 
equations:
\begin{equation}
{dp_i\over dt}=-{\partial {\cal H}\over \partial q_i}
\label{eq:b}
\end{equation}
\begin{equation}
{dq_i\over dt}={\partial {\cal H}\over \partial p_i}
\label{eq:c}
\end{equation}
For our specific dynamic problems in a circular accelerator it is convenient
to chose curvilinear coordinates instead of Cartesian ones for us to describe
the trajectory of a particle near an {\it a priori} known closed orbit.
The Hamiltonian in the new system, $(x, s, y)$ (where $x, s$, and $y$
denote the coordinates in the Frenet-Serret normal, tangent, 
and binormal triorthogonal and right-handed coordinate system), 
is given by \cite{1}$^-$\cite{7}:
\begin{equation}
H_t(q,p,t)=e\Phi +c\left((p_x-eA_x)^2+(p_y-eA_y)^2+
\left({p_s-eA_s\over 1+x/\rho}\right)^2+m_0c^2\right)^{1/2}
\label{eq:d}
\end{equation}
where $\rho$ is the radius of curvature and the torsion of the closed orbit
is everywhere zero. Since it is useful to use the variable, $s$, as the 
independent variable rather than time, $t$, 
one gets the new Hamiltonian by using a simple canonical transformation:
\begin{equation}
H_s=-eA_s-(1+x/\rho)\left({1\over c^2}(E^2-m_0^2c^4)-(p_x-eA_x)^2
-(p_y-eA_y)^2\right)^{1/2}-e\Phi   
\label{eq:e}
\end{equation}
Noting that the term ${1\over c^2}(E^2-m_0^2c^4)$ in eq. \ref{eq:e} is
equal to $P^2$ with $P$ being the total mechanical momentum of the particle,
by using another trivial canonical transformation:
\begin{equation}
{\bar q}=q,\space {\bar s}=s,\space {\bar p}_{x,y}={p_{x,y}\over P_0},
\space H={H_s\over P_0}
\label{eq:f}
\end{equation}
one gets another Hamiltonian:
\begin{equation}
H=-{eA_s\over P_0}-(1+x/\rho)\left({P\over P_0}-({\bar p}_x-{eA_x\over P_0})^2
-({\bar p}_y-{eA_y\over P_0})^2\right)^{1/2}-{e\Phi \over P_0}
\label{eq:g}
\end{equation}
where $P_0$ is the mechanical momentum of the reference particle, and 
$P=P_0+\Delta P$.
Inserting $eA_s/P_0$ in eq. \ref{eq:g} by:
\begin{equation}
{eA_s\over P_0}=-{B_yx^2\over 2\rho^2 B_0}-
{1\over B_0\rho}\sum_{n=1}^{\infty}{1\over n!}
{\partial^{n-1}B_y\over \partial x^{n-1}}\vert_{x=0,y=0}
(x+iy)^n
\label{eq:h}
\end{equation}
one gets finally the Hamiltonian which serves as the 
starting point of most of the dynamical problems concerning circular accelerators:
$$H={x^2B_y\vert_{x=0,y=0}\over 2\rho^2 B_0}+
{1\over B_0\rho}\sum_{n=1}^{\infty}
{1\over n!}{\partial^{n-1}B_y\over \partial x^{n-1}}
\vert_{x=0,y=0}
(x+iy)^n$$
\begin{equation}
-(1+x/\rho)\left(1+{\Delta P\over P_0}-
\left({\bar p}_x-{eA_x\over P_0}\right)^2
-\left({\bar p}_y-{eA_y\over P_0}\right)^2\right)^{1/2}-{e\Phi \over P_0}
\label{eq:i}
\end{equation}
where $B_0$ is the bending magnetic field on the orbit of the 
reference particle, and $B_y$ in general is a complex variable.  
\par
\section{Analytical formulae for dynamic apertures}
To start with we consider the linear horizontal 
motion of the reference particle (no energy deviation) in the
horizontal plane (y=0)  
assuming that the magnetic field is only transverse ($A_x=A_y=0$) and has no
skew fields, and $\Phi$ is a constant. The Hamiltonian can be 
simplified as 
\begin{equation}
H={p^2\over 2}+{K(s)\over 2}x^2
\label{eq:1}
\end{equation}
where $x$ denotes normal plane coordinate, $p=dx/ds$,
and $K(s)$ is a periodic function satisfying the relation
\begin{equation}
K(s)=K(s+L)
\label{eq:1a}
\end{equation}
where $L$ is the circumference of the ring. The solution of the deviation,
$x$, is found to be 
\begin{equation}
x=\sqrt{\epsilon_x\beta_x (s)}\cos(\phi (s)+\phi_0)
\label{eq:2}
\end{equation}
where
\begin{equation}
\phi (s)=\int_0^s{ds\over \beta_x(s)}
\label{eq:3}
\end{equation}
As an essential step towards further discussion on the motions under nonlinear
perturbation forces, 
we introduce action-angle variables and the Hamiltonian expressed in these
new variables:
\begin{equation}
\Psi=\int_0^s{ds'\over \beta_x(s')}+\phi_0
\label{eq:5}
\end{equation}
\begin{equation}
J={\epsilon_x \over 2}
={1\over 2\beta_x (s)}
\left(x^2+\left(\beta_x (s)x'-{\beta_x'x\over 2}\right)^2\right)
\label{eq:6}
\end{equation}
\begin{equation}
H(J,\Psi)={J\over \beta_x(s)}
\label{eq:4}
\end{equation}
Since $H(J,\Psi)=J/ \beta_x(s)$ is still a function of the independent
variable, $s$, we will make another canonical transformation to freeze the
new Hamiltonian:
\begin{equation}
\Psi_1=\Psi+{2\pi \nu \over L}-\int_0^s{ds'\over \beta_x(s')} 
\label{eq:8}
\end{equation}
\begin{equation}
J_1=J
\label{eq:9}
\end{equation}
\begin{equation}
H_1={2\pi \nu \over L}J_1
\label{eq:10}
\end{equation}
Before going on further, 
let's remember the relation between the last action-angle
variables and the particle deviation $x$:
\begin{equation}
x=\sqrt{2J_1\beta_x(s)}\cos\left(\Psi_1-{2\pi \nu\over L}s
+\int_0^s{ds'\over \beta_x(s')}\right)
\label{eq:11}
\end{equation}
Being well prepared, we start our journey to find out the limitations of the
nonlinear forces on the stability of the particle's motion. 
To facilitate the analytical treatment of this complicated problem
we consider at this stage only sextupoles and octupoles (no skew terms) and
assume that the contributions from the sextupoles and octupoles in a ring
can be made equivalent to a point sextupole and a point octupole. The perturbed
one dimensional Hamiltonian can thus be expressed:
\begin{equation}
H={p^2\over 2}+{K(s)\over 2}x^2+{1\over 3!B\rho}{\partial^2B_z\over 
\partial x^2}x^3L\sum_{k=-\infty}^{\infty}\delta (s-kL)
+{1\over 4!B\rho}{\partial^3B_z\over
\partial x^3}x^4L\sum_{k=-\infty}^{\infty}\delta (s-kL)
\label{eq:12}
\end{equation}
Representing eq. \ref{eq:12} by action-angle variables ($J_1$ and $\Psi_1$),
and using 
\begin{equation}
B_z=B_0(1+xb_1+x^2b_2+x^3b_3)
\label{eq:12a}
\end{equation}
one has
$$H={2\pi \nu \over L}J_1+{(2J_1\beta_x (s_1))^{3/2}\over 3\rho}
b_2L\cos^3 \Psi_1\sum_{k=-\infty}^{\infty}\delta (s-kL)$$
\begin{equation}
+{(J_1\beta_x (s_2))^2\over \rho}b_3L\cos^4\Psi_1
\sum_{k=-\infty}^{\infty}\delta (s-kL)
\label{eq:13}
\end{equation}
where $s_1$ and $s_2$ are just used to differentiate the locations of the
sextupole and the octupole perturbations. 
By virtue of Hamiltonian one gets the differential equations for $\Psi_1$
and $J_1$
\begin{equation}
{dJ_1\over ds}=-{\partial H_1\over \partial \Psi_1}
\label{eq:15}
\end{equation}
\begin{equation}
{d\Psi_1\over ds}={\partial H_1\over \partial J_1}
\label{eq:14}
\end{equation}  
$${dJ_1\over ds}=-{(2J_1\beta_x (s_1))^{3/2}\over 3\rho}
b_2L{d\cos^3 \Psi_1\over d \Psi_1}\sum_{k=-\infty}^{\infty}\delta (s-kL)$$
\begin{equation}
-{(J_1\beta_x (s_2))^2\over \rho}b_3L{d\cos^4\Psi_1\over d\Psi_1}
\sum_{k=-\infty}^{\infty}\delta (s-kL)
\label{eq:16}
\end{equation}
$${d\Psi_1\over ds}={2\pi \nu \over L}+{\sqrt{2}J_1^{1/2}\beta_x (s_1)^{3/2}
\over \rho}
b_2L\cos^3 \Psi_1\sum_{k=-\infty}^{\infty}\delta (s-kL)$$
\begin{equation}
+{2\beta_x^2(s_2)\over \rho}J_1b_3L\cos^4\Psi_1
\sum_{k=-\infty}^{\infty}\delta (s-kL)
\label{eq:17}
\end{equation}
We now change these differential equations to the difference
equations which are {\it suitable} to analyse the possibilities of the onset of 
stochasticity \cite{9}\cite{10}. 
Since the perturbations have a natural periodicity of $L$ we will sample
the dynamic quantities at a sequence of $s_i$ with constant interval $L$
assuming that the characteristic time between two consecutive adiabatic
invariance breakdown intervals is shorter than $L/c$.
The differential equations in eqs. \ref{eq:16} and \ref{eq:17} are reduced to
\begin{equation}
\overline{J_1}=\overline{J_1}(\Psi_1,J_1)
\label{eq:19}
\end{equation} 
\begin{equation}
\overline{\Psi_1}=\overline{\Psi_1}(\Psi_1,J_1)
\label{eq:18}
\end{equation}
where the bar stands for the next sampled value after the corresponding
unbarred previous value.  
\begin{equation}
\overline{J_1}=J_1-{(2J_1\beta_x (s_1))^{3/2}\over 3\rho}
b_2L{d\cos^3 \Psi_1\over d \Psi_1}
-{(J_1\beta_x (s_2))^2\over \rho}b_3L{d\cos^4\Psi_1\over d\Psi_1}
\label{eq:20}
\end{equation}
\begin{equation}
\overline{\Psi_1}=\Psi_1+{2\pi \nu }+{\sqrt{2}\beta_x (s_1)^{3/2}
\overline{J_1}^{1/2}
\over \rho}b_2L\cos^3 \Psi_1
+{2\beta_x (s_2)^2\over \rho}\overline{J_1}b_3L\cos^4\Psi_1
\label{eq:21}
\end{equation}
Eqs. \ref{eq:20} and \ref{eq:21} are the basic difference equations
to study the nonlinear resonance and the onset of stochasticities considering
sextupole and octupole perturbations.
By using trigonometric relation
\begin{equation}
\cos^m\theta \cos n\theta=2^{-m}\sum_{r=0}^m{m!\over (m-r)!r!}\cos (n-m+2r)
\theta
\label{eq:22}
\end{equation}
one has
\begin{equation}
\cos^3 \theta ={2\over 2^3}(\cos3\theta+3\cos\theta)
\label{eq:23}
\end{equation}
\begin{equation}
\cos^4 \theta ={1\over 2^4}(2\cos4\theta+8\cos2\theta +{4!\over ((4/2)!)^2})
\label{eq:24}
\end{equation}
If the tune $\nu$ is far from the resonance lines $\nu=m/n$, where
$m$ and $n$ are integers, the
invariant tori of the unperturbed motion are preserved under the presence
of the small perturbations by virtue of the 
Kolmogorov-Arnold-Moser (KAM) theorem. If, however, 
$\nu$ is close to the above mentioned 
resonance line, under some conditions the KAM invariant tori can be broken.
\par
Consider first the case where there is only one sextupole located at
$s=s_1$ with $\beta_x(s_1)$.
Taking the third order resonance, $m/3$, 
for example, we keep only the sinusoidal 
function with phase $3\Psi_1$ in eq. \ref{eq:20} and the
dominant phase independent nonlinear term in eq. \ref{eq:21}, 
and as the result, 
eqs. \ref{eq:20} and \ref{eq:21} become
\begin{equation}
\overline{J_1}=J_1+A\sin 3\Psi_1
\label{eq:25a}
\end{equation}
\begin{equation}
\overline{\Psi_1}=\Psi_1
+B\overline{J_1}
\label{eq:26a}
\end{equation}
with
\begin{equation}
A={(2J_1\beta_x (s_1))^{3/2}\over 4}\left({
b_2L\over \rho}\right) 
\label{eq:25}
\end{equation}
\begin{equation}
B=\sqrt{2}\beta_x (s_1)^{3/2}J_1^{-1/2}\left({b_2L\over \rho}\right)   
\label{eq:26}
\end{equation}
where we have dropped the constant phase in eq. \ref{eq:21}
and take the maximum value of $\cos^3({\Psi_1})$, 1. It is helpful
to transform eqs. \ref{eq:25} and \ref{eq:26} into the form so-called 
{\it standard mapping} \cite{10} expressed as
\begin{equation}
\overline{I}=I+K_0\sin \theta
\label{eq:27}
\end{equation}
\begin{equation}
\overline{\theta}=\theta+\overline{I}
\label{eq:28}
\end{equation}
with $\theta=3\Psi$, $I=3BJ_1$ and $K_0=3AB$. By virtue of the Chirikov
criterion \cite{10} 
it is known that when $\vert K_0\vert \geq  
0.97164$ \cite{11} resonance overlapping occurs which results in
particles' stochastic motions and diffusion processes. Therefore, 
\begin{equation}
\vert K_0\vert\ \leq 1
\label{eq:29}
\end{equation}
can be taken as a natural 
criterion for the determination of the dynamic aperture of the machine.
Putting eqs. \ref{eq:25} and \ref{eq:26} into eq. \ref{eq:29}, one gets   
\begin{equation}
\vert K_0\vert =3J_1\beta_x(s_1)^3\left({\vert b_2\vert L
\over \rho }\right)^2\leq 1
\label{eq:30}
\end{equation}  
and consequently, one finds maximum $J_1$ corresponding to a $m/3$ resonance
\begin{equation}
J_1\leq J_{max,sext}={1\over 3\beta_x(s_1)^3}\left({\rho 
\over \vert b_2\vert L}\right)^2
\label{eq:31}
\end{equation}
The dynamic aperture of the machine is therefore
\begin{equation}
A_{dyna,sext}=\sqrt{2J_{max,sext}\beta_{x}(s)}
={\sqrt{2\beta_x(s)}\over \sqrt{3}\beta_x(s_1)^{3/2}}\left({\rho 
\over \vert b_2\vert L}
\right)
\label{eq:32}
\end{equation}
Eq. \ref{eq:32} gives the dynamic aperture of a sextuple
strength determined case. The reader can confirm that if we keep 
$\sin(\Psi_1)$ term instead of $\sin(3\Psi_1)$ in eq. \ref{eq:25a}, one arrives
at the same expression for $A_{dyna,sext}$ as expressed in eq. \ref{eq:32}.
\par
Secondly, we consider the case of single octupole located at $s=s_2$ with
$\beta_x(s_2)$.
Taking the forth order resonance, $m/4$, 
for example, we keep only the sinusoidal 
function with phase $4\Psi_1$ in eq. \ref{eq:20} and the
dominant phase independent nonlinear term in eq. \ref{eq:21}, 
and as the result, we have
eqs. \ref{eq:20} and \ref{eq:21} reduced to
\begin{equation}
\overline{J_1}=J_1+A\sin 4\Psi_1
\label{eq:25aa}
\end{equation}
\begin{equation}
\overline{\Psi_1}=\Psi_1
+B\overline{J_1}
\label{eq:26aa}
\end{equation}
with
\begin{equation}
A={(J_1\beta_x (s_2))^{2}\over 2}\left({
b_3L\over \rho}\right) 
\label{eq:25pp}
\end{equation}
\begin{equation}
B={2\beta_x (s_2)^2}\left({b_3L\over \rho}\right)   
\label{eq:26pp}
\end{equation}
where we have dropped the constant phase in eq. \ref{eq:21}
and take the maximum value of $\cos^4({\Psi_1})$, 1.
By using Chirikov criterion, one gets
\begin{equation}
J_1\leq J_{max,oct}={1\over 2\beta_x(s_2)^2}\left({\rho 
\over \vert b_3\vert L}\right)
\label{eq:33}
\end{equation}
and the corresponding dynamic aperture:
\begin{equation}
A_{dyna,oct}=\sqrt{2J_{max,oct}\beta_{x}(s)}
={\sqrt{\beta_x(s)}\over \beta_x(s_2)}\sqrt{\rho \over \vert b_3\vert L}
\label{eq:34}
\end{equation}
Thirdly, without repeating, we give directly the dynamic aperture due to a
decapole located at $s=s_3$:
\begin{equation}
A_{dyna,deca}
=\sqrt{2\beta_x(s)}\left({1\over 5\beta_x^5(s_3)}\right)^{1/6}
\left({\rho \over \vert b_4\vert L}
\right)^{1/3}
\label{eq:34aa}
\end{equation}
where $b_4$ is the coefficient of the decapole strength.
Finally, we give the general expression of the dynamic aperture in the
horizontal plane ($z=0$) of a single $2m$ ($m \geq 3$) pole component:
\begin{equation}
A_{dyna,2m}
=\sqrt{2\beta_x(s)}\left({1\over m\beta_x^m(s(2m))}\right)^{1\over 2(m-2)}
\left({\rho \over \vert b_{m-1}\vert L}
\right)^{1/(m-2)}
\label{eq:34aabb}
\end{equation}
where $s(2m)$ is the location of this multipole.
Eq. \ref{eq:34aabb} gives us useful scaling laws, such as
$A_{dyna,2m}\propto \left({\rho \over \vert b_{m-1}\vert L}
\right)^{1/(m-2)}$, and 
$A_{dyna,2m}\propto \left({1\over \beta_x^m(s(2m))}\right)^{1\over 2(m-2)}$.
\par
If there is more than one
nonlinear component, how can one 
estimate their collective effect ? Fortunately, one can distinguish two cases:
\begin{itemize}
\item[1)] If the components are {\it independent}, i.e. there are no special
phase and amplitude relations between them, the 
total dynamic aperture
can be calculated as:
\end{itemize}
\begin{equation}
A_{dyna,total}={1\over \sqrt{\sum_i{1\over A_{dyna,sext,i}^2}
+\sum_j{1\over A_{dyna,oct,j}^2}+\sum_k{1\over A_{dyna,deca,k}^2}+\cdot \cdot \cdot}}
\label{eq:47}
\end{equation}
\begin{itemize}
\item[2)] If the nonlinear components are {\it dependent}, 
i.e. there are special phase and amplitude relations between them
(for example, in reality, one use some additional sextupoles to 
cancel the nonlinear effects of the sextupoles used to make chromaticity
corrections), there is no general formula as eq. \ref{eq:47} to apply.
\end{itemize}   
\par
In the above discussion we have restricted us to the case where 
particles are moving in the horizontal plane, and the one dimensional
dynamic aperture
formulae expressed in eqs. \ref{eq:32}, \ref{eq:34}, \ref{eq:34aa}
and \ref{eq:47}, are the maximum stable horizontal excursion ranges with
the vertical displacement $y=0$. In the following we will show briefly how
to estimate the dynamic aperture in 2 dimensions when there is coupling 
between the horizontal and vertical planes. Now we consider the case where
only one sextupole is located at $s=s_1$, and we have the corresponding
Hamiltonian expressed as follows:
\begin{equation}
H={p_x^2\over 2}+{K_x(s)\over 2}x^2
+{p_y^2\over 2}+{K_y(s)\over 2}y^2
+{1\over 3!B\rho}{\partial^2B_z\over 
\partial x^2}(x^3-3xy^2)L\sum_{k=-\infty}^{\infty}\delta (s-kL)
\label{eq:12bb}
\end{equation}
Generally speaking, there exists no universal criterion to determine the start up 
of stochastic motions
in 2D. Fortunately, in our specific case, we find out the similarity
between the Hamiltonian expressed in eq. \ref{eq:12bb} and that of 
the H\'enon and Heiles problem which has been much studied in literature
\cite{11a}.
The H\'enon and Heiles problem's Hamiltonian is given by
\begin{equation}
H_{H\&H}={1\over 2}\left(x^2+p_x^2+y^2+p_y^2+2y^2x-{2\over 3}x^3\right)
\label{eq:12cc}
\end{equation}
when $H_{H\&H} > 1/6$ the motion becomes unstable.
The intuition we get from this conclusion is that there should exist 
a similar criterion for our problem, i.e. to have stable 2D motion one 
should have $H\leq H_{max}$. Note that 
$K_{x}(s)$ and $K_y(s)$ in eq. \ref{eq:12bb} are equal to unity in 
the H\'enon and Heiles problem's Hamiltonian.
The previous one dimensional result  
helps us now to find $H_{max}$. When $y=0$ one has 
$H_{max}\propto A^2_{dyna,sext,x}$, since $x\leq A_{dyna,sext,x}$.
When $y\neq 0$, the crossing terms in eqs. \ref{eq:12bb} and 
\ref{eq:12cc} will play the role of exchanging energy between the two planes,
and for a given set of $x$ and $y$ the total energy of the coupled system 
can not exceed $H_{max}$. If we define $A_{dyna,sext,y}$ 
is the dynamic aperture in $y$-plane, one has: 
\begin{equation}
\beta_x(s_1)A_{dyna,sext,x}^2=\beta_y(s_1)A^2_{dyna,sext,y}+\beta_x(s_1)x^2
\label{eq:12ggg}
\end{equation}
or:
\begin{equation}
A_{dyna,sext,y}=\sqrt{{\beta_x(s_1)\over \beta_y(s_1)}(A^2_{dyna,sext,x}-x^2)}
\label{eq:12gg}
\end{equation}
where $\beta_y(s_1)$ is the vertical beta function where the sextupole
is located and $A_{dyna,sext,x}$ is given by eq. \ref{eq:32}. 
Different from eq. \ref{eq:34aabb},
the derivation of eq. \ref{eq:12gg} is quite intuitive, hinted by 
the H\'enon and Heiles problem which has been studied numerically instead
of analytically in literature.
From
eq. \ref{eq:12gg} one understands that the difference between $A_{dyna,sext,y}$
and $A_{dyna,sext,x}$ comes from $\sqrt{\beta_x(s_1)/ \beta_y(s_1)}$.
If there are many sextupoles in a ring one usually has 
$A_{dyna,sext,x}\approx A_{dyna,sext,y}$ since $\beta_x(s_i)$ will not
be always larger or smaller than $\beta_y(s_i)$.
\par
\section{Comparison with simulation results}
\begin{figure}[h]
\vskip 1.5 true cm
\vspace{6.0cm}
\includegraphics{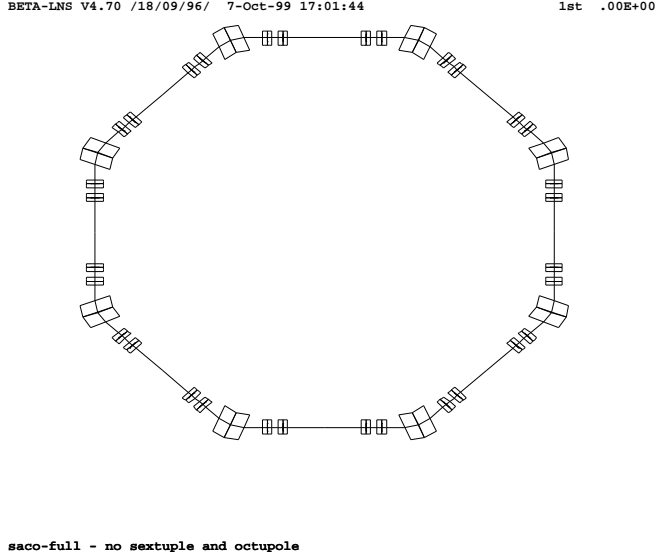}
\vskip 0.5 true cm
\caption{The schematic layout of Super-ACO.
\label{fig101}}
\end{figure}

To verify the validity of eqs. \ref{eq:32}, \ref{eq:34}, \ref{eq:47},
and \ref{eq:12gg}, we compare
the dynamic apertures of some special cases 
by using these analytical formulae with a computer code called BETA
\cite{12}. 
Taking the lattice of Super-ACO as an example, we show the schematic layout
of the machine in \linebreak Fig. \ref{fig101}.\\ 

The horizontal beta function distribution and the working point in the third
order tune diagram are given in Figs. \ref{fig102} and \ref{fig103},
respectively. From Fig. \ref{fig102} one finds that the horizontal beta
function at the beginning and the end of the one turn mapping is 5.6 m.
In the following numerical simulations the dynamic apertures correspond
to $\beta_x(s)=\beta_x(0)=5.6$ m and it will also be used in the analytical formulae. 
Defining the sextupole, octupole, and decapole
 strengths $S=b_2L/\rho$ (1/m$^2$),
$O=b_3L/\rho$ (1/m$^3$), $D=b_4L/\rho$ (1/m$^4)$, respectively, we make now a rather systematic comparison.
\vspace{5mm}
\begin{itemize}
\item[1)] A sextupole is located at $s=s_1$ with $S(s_1)=1$ and $\beta_x(s_1)=13.6$ m,
and its influence on the horizontal dynamic aperture is illustrated in 
Figs. \ref{fig104} and \ref{fig105}. From eq. \ref{eq:32} one gets
analytically that $A_{dyna,sext}=0.0385$ m compared with the numerical
value of 0.04 m.
\item[2)] An octupole is located at $s=s_1$ with $O(s_1)=10$ and 
$\beta_x(s_1)=13.6$ m,
and its influence on the horizontal dynamic aperture is illustrated in 
Figs. \ref{fig106} and \ref{fig107}. From eq. \ref{eq:34} one gets
analytically that $A_{dyna,oct}=0.055$ m compared with the numerical
value of 0.054 m.
\item[3)] The validity 
of eq. \ref{eq:34aa} has been checked also. If a decapole with strength
$D=1000$ is located at
$s=s_1$, the dynamic aperture in horizontal plane is shown in 
Figs. \ref{fig114aa} and  \ref{fig115aa}. From eq. \ref{eq:34aa} one gets
$A_{dyna,deca}=0.022$ m compared with the numerical
value of 0.024 m.
\item[4)] A sextupole of $S=2$ and a octupole of $O=62$ are located at $s=s_1$ and 
$\beta_x(s_1)=13.6$ m, and their combined influence on the horizontal dynamic 
aperture is shown in Figs. \ref{fig108} and \ref{fig109}.
From eq. \ref{eq:47} one gets $A_{dyna,total}=0.0145$ m compared with the numerical value of 0.016 m.
\newpage
\begin{figure}[h]
\vskip 2.5 true cm
\vspace{6.0cm}
\includegraphics{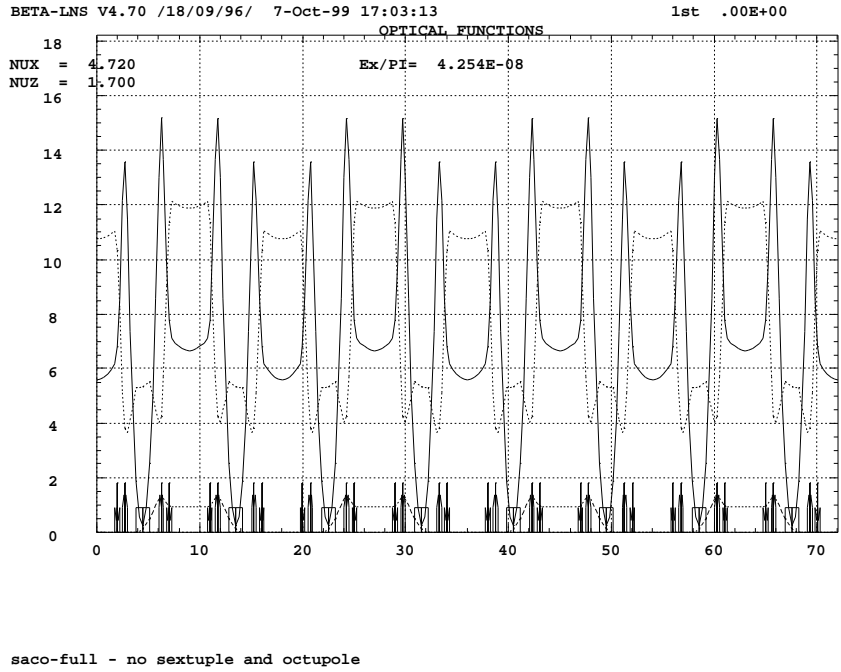}
\vskip 0.5 true cm
\caption{The horizontal beta function distribution of Super-ACO 
($\beta_x(0)=5.6$ m).
\label{fig102}}
\end{figure}
\vspace*{50mm}
\begin{figure}[h]
\vskip 0.5 true cm
\vspace{6.cm}
\includegraphics{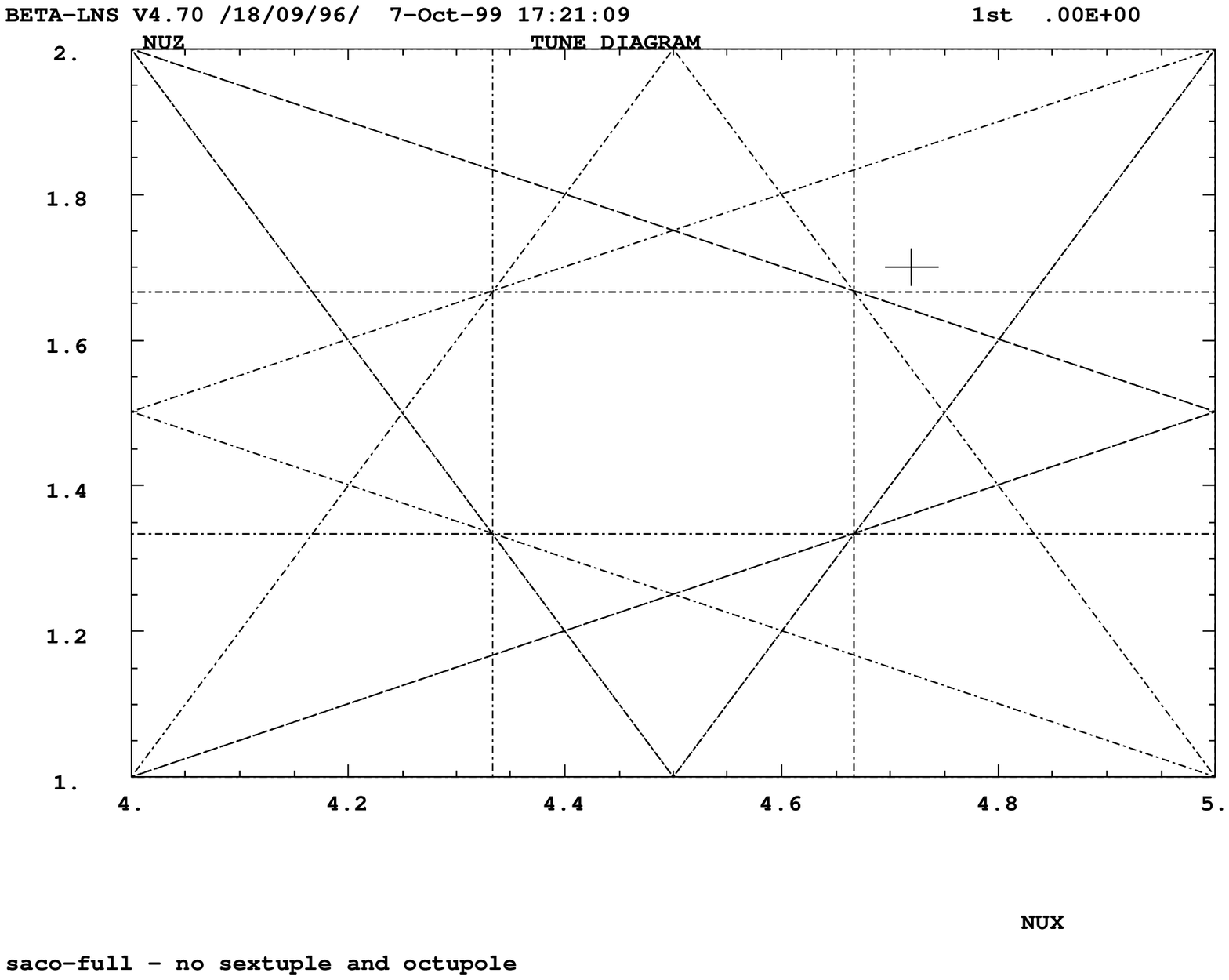}
\vskip 0. true cm
\caption{The tune diagram of the third order of Super-ACO, where the cross
indicates the working point of the machine.
\label{fig103}}
\end{figure}
\newpage

\newpage
\begin{figure}[h]
\vskip 2.5 true cm
\vspace{6.0cm}
\includegraphics{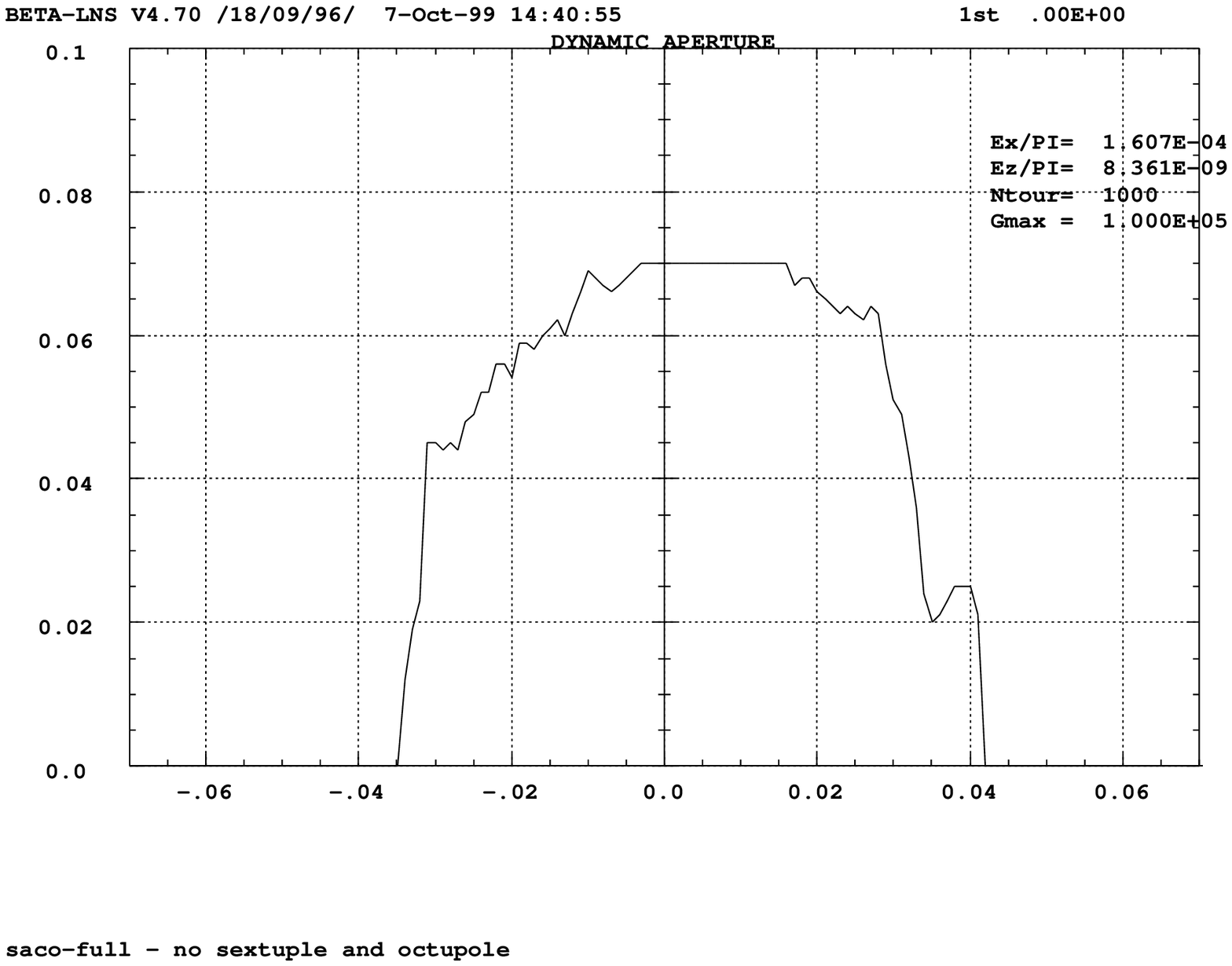}

\vskip 1.5 true cm
\caption{The dynamic aperture plot ($S(s_1)=1$ and $\beta_x(s_1)=13.6$ m).
\label{fig104}}
\end{figure}
\vspace*{40mm}

\begin{figure}[h]
\vskip 0. true cm
\vspace{6.0cm}
\includegraphics{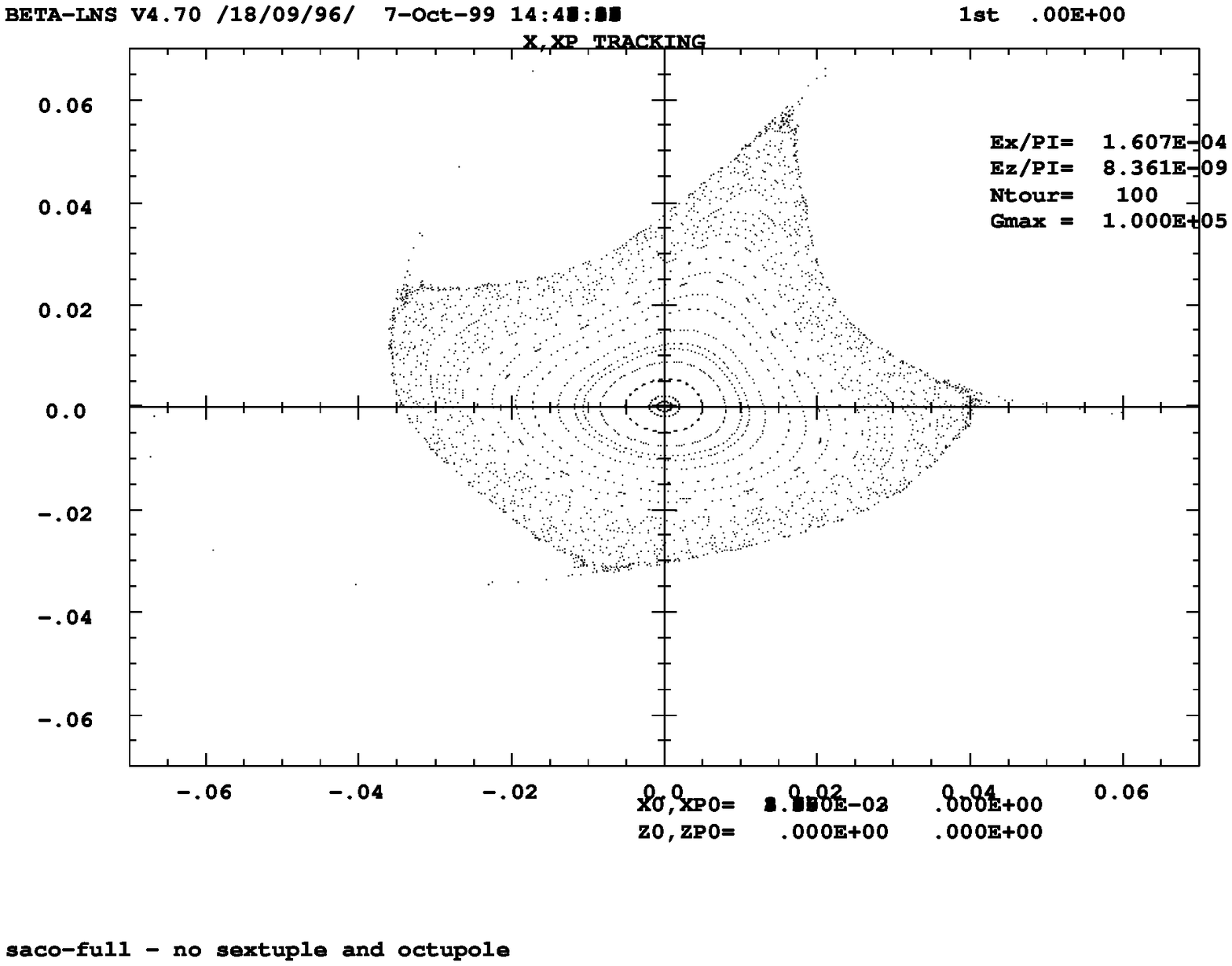}
\vskip 0. true cm
\caption{The horizontal phase space ($S(s_1)=1$ and $\beta_x(s_1)=13.6$ m).
\label{fig105}}
\end{figure}
\newpage
\begin{figure}[h]
\vskip 2.5 true cm
\vspace{6.0cm}
\includegraphics{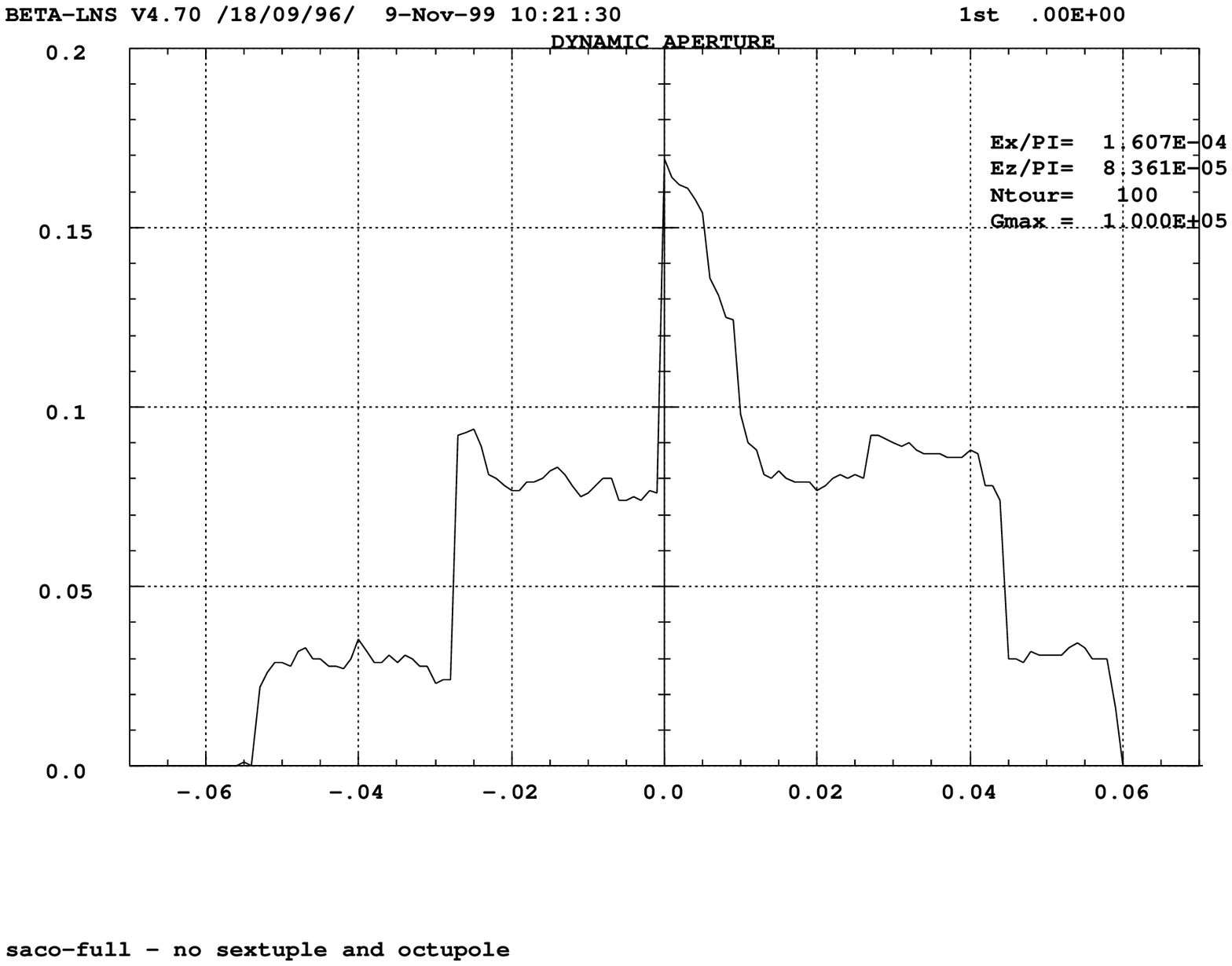}
\vskip 0. true cm
\caption{The dynamic aperture plot ($O(s_1)=10$ and $\beta_x(s_1)=13.6$ m).
\label{fig106}}
\end{figure}
\vspace*{50mm}
\begin{figure}[h]
\vskip 1. true cm
\vspace{6.0cm}
\includegraphics{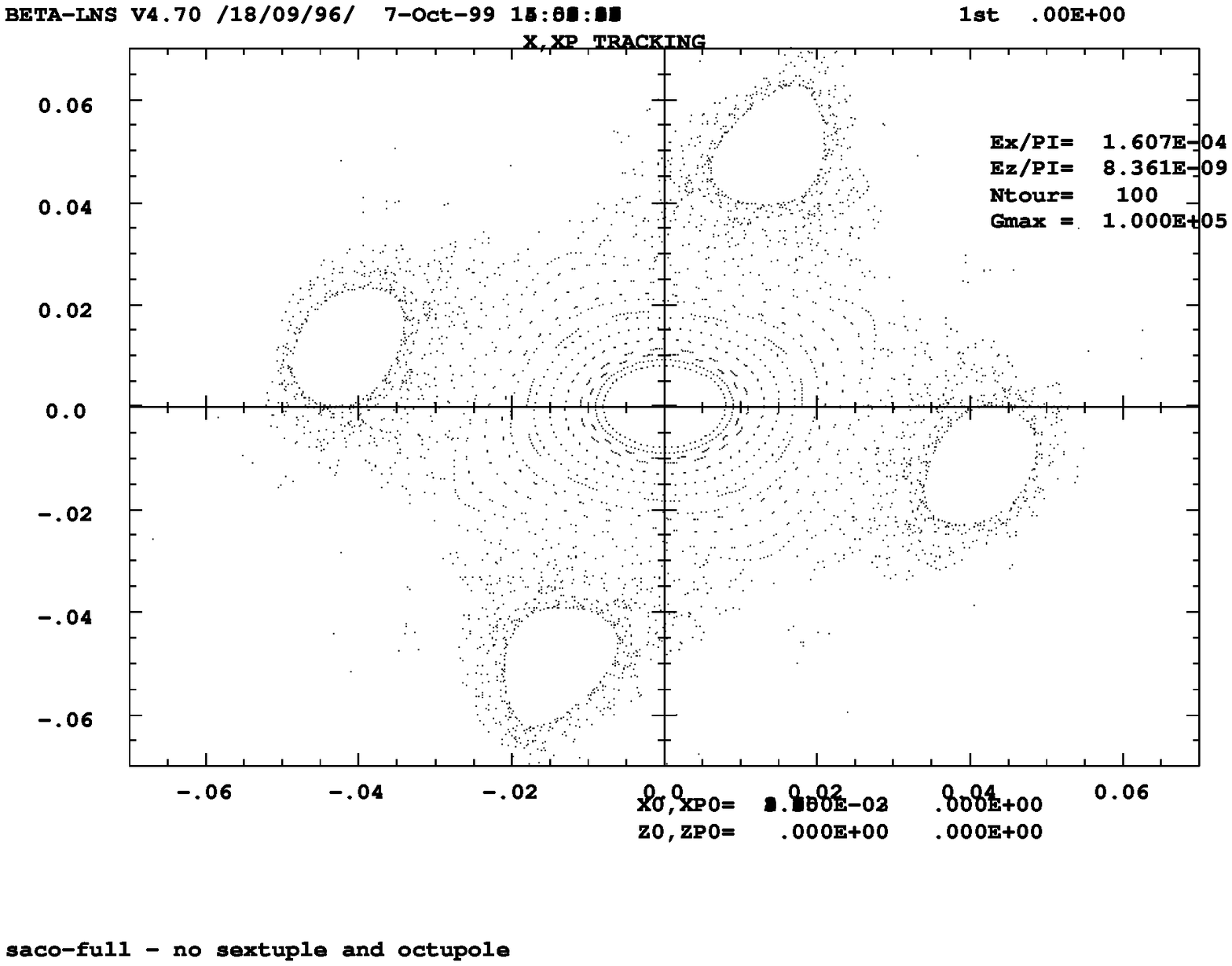}
\vskip 0. true cm
\caption{The horizontal phase space ($O(s_1)=10$ and $\beta_x(s_1)=13.6$ m).
\label{fig107}}
\end{figure}
\newpage
\begin{figure}[h]
\vskip 2.6 true cm
\vspace{6.0cm}
\includegraphics{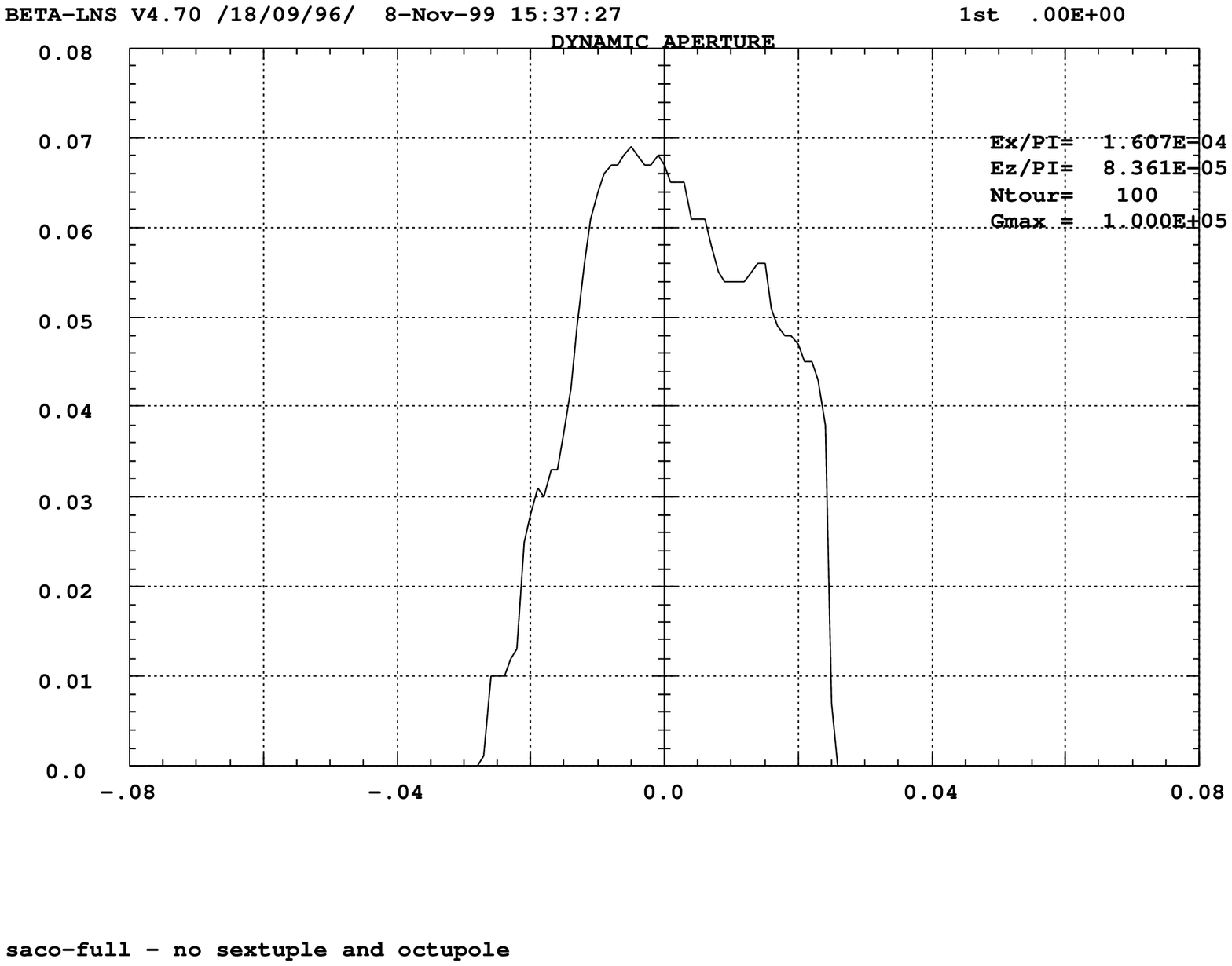}
\vskip 0. true cm
\caption{The dynamic aperture plot ($D(s_1)=1000$ and $\beta_x(s_1)=13.6$ m).
\label{fig114aa}}
\end{figure}
\vspace*{40mm}
\begin{figure}[h]
\vskip 1.5 true cm
\vspace{6.0cm}
\includegraphics{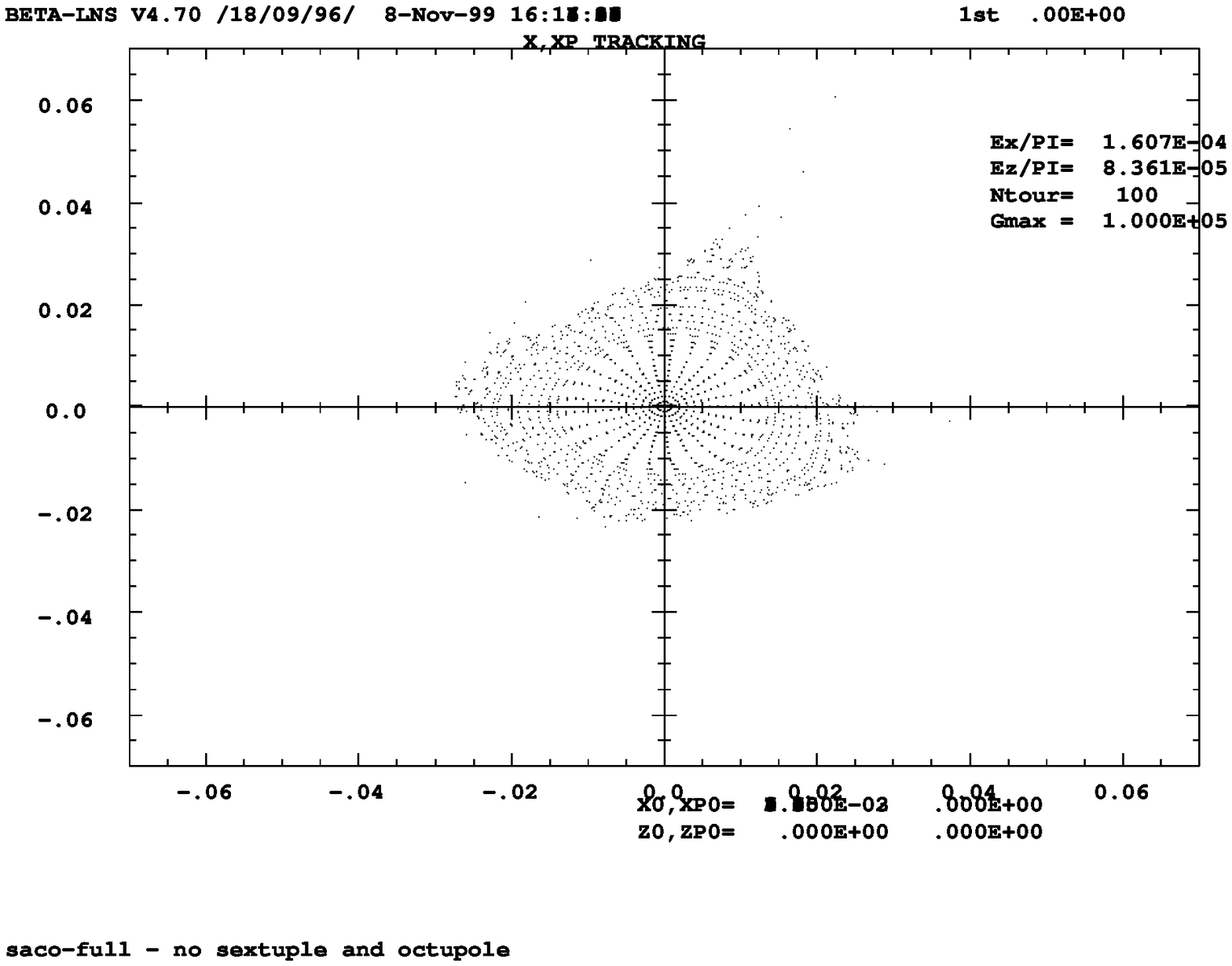}
\vskip 0. true cm
\caption{The horizontal phase space ($D(s_1)=1000$ and
$\beta_x(s_1)=13.6$ m). 
\label{fig115aa}}
\end{figure}
\newpage
\begin{figure}[h]
\vskip 2.5 true cm
\vspace{6.0cm}
\includegraphics{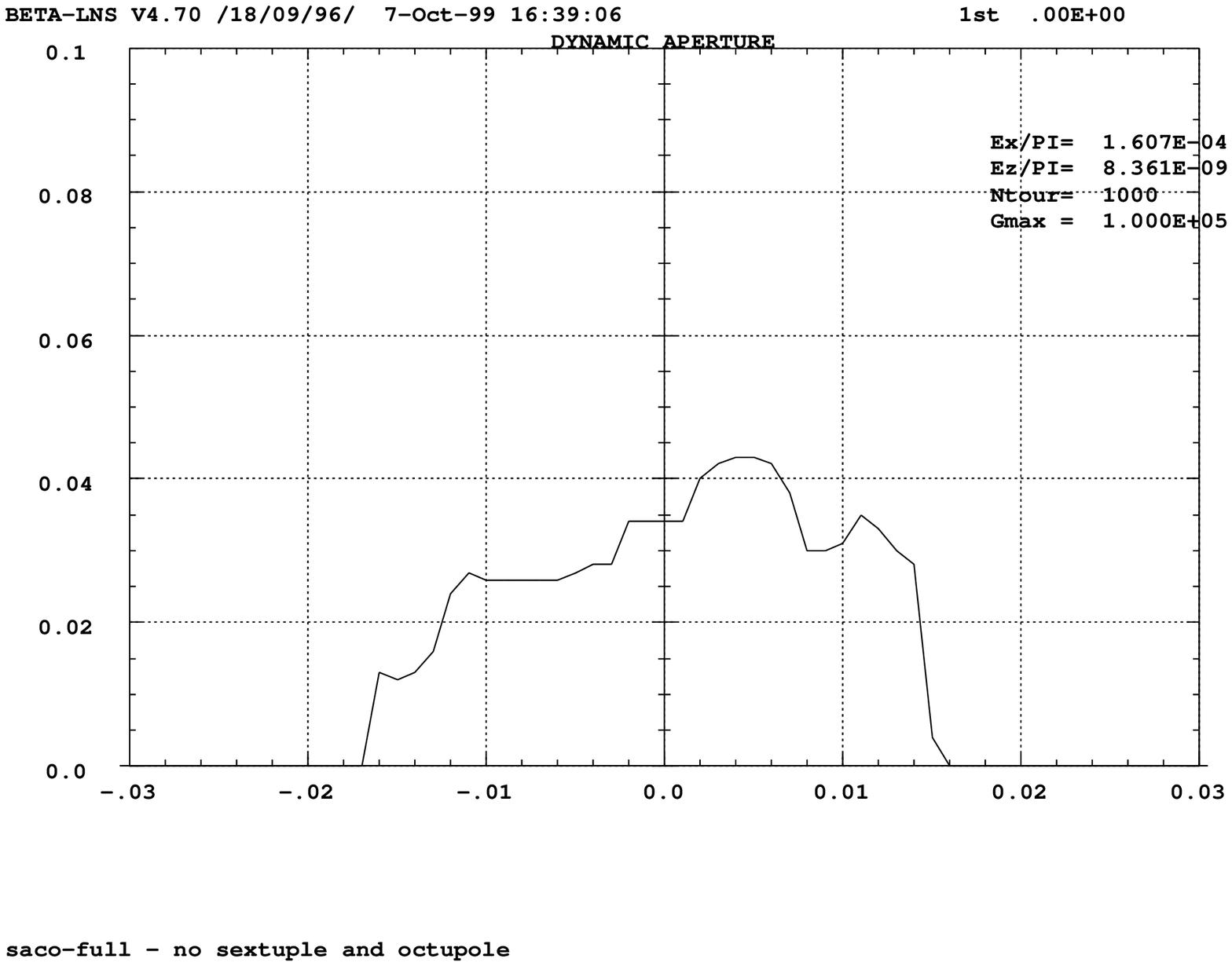}
\vskip 0. true cm
\caption{The dynamic aperture plot ($S(s_1)=2$, $O(s_1)=62$, 
and $\beta_x(s_1)=13.6$ m).
\label{fig108}}
\end{figure}
\vspace*{40mm}
\begin{figure}[h]
\vskip 0. true cm
\vspace{7.0cm}
\includegraphics{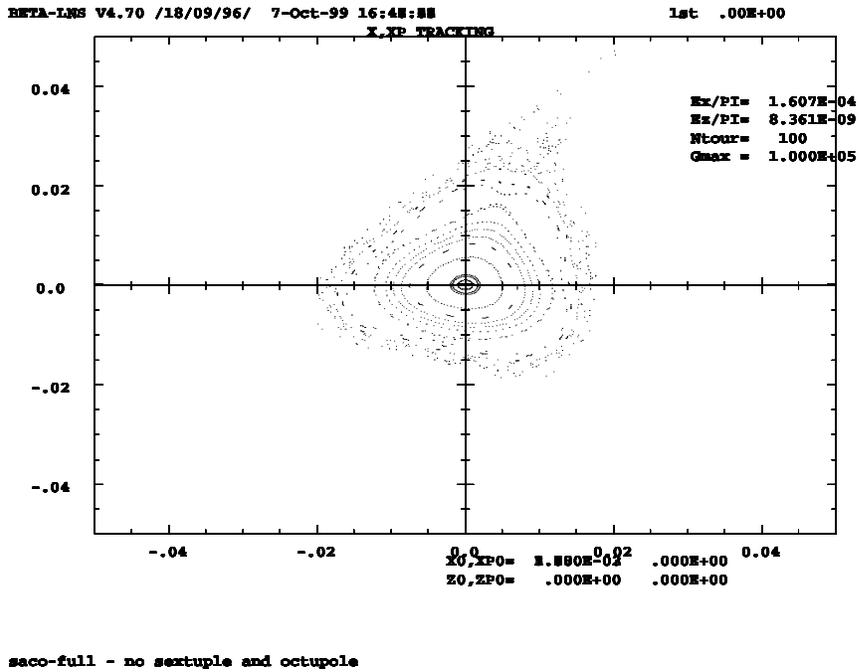}
\vskip 0. true cm
\caption{The horizontal phase space ($S(s_1)=2$, $O(s_1)=62$, 
and $\beta_x(s_1)=13.6$ m). 
\label{fig109}}
\end{figure}
\newpage
\item[5)] A sextupole of $S=2$ and a octupole of $O=62$ are
 located at $s=s_1$, $\beta_x(s_1)=13.6$ m, and $s=s_2$, 
$\beta_x(s_2)=15.18$ m, respectively, and their combined influence on the horizontal dynamic aperture is shown in Figs. \ref{fig110} and \ref{fig111}.
From eq. \ref{eq:47} one gets $A_{dyna,total}=0.0138$ m
compared with the numerical
value of 0.0135 m.
\item[6)] Four sextupoles of $S=2$ are located at $s=s_{1,2,3,4}$ with 
$\beta_x(s_1)=13.6$ m, $\beta_x(s_2)=15.18$ m, $\beta_x(s_3)=7.8$ m,
and $\beta_x(s_4)=6.8$ m, respectively, and their combined influence on the horizontal dynamic aperture is given in Figs. \ref{fig110a} and \ref{fig111a}.
From eq. \ref{eq:47} one obtained the analytical value 
$A_{dyna,total}=0.012$ m.
\item[7)] We show how the dynamic apertures depend on the strengths of
sextupole and octupole. 
Fig. \ref{fig116} shows the comparison between the 
analytical (solid line) and the numerical (dotted) results for a sextupole 
located at $s=s_1$ in the machine shown in Fig. \ref{fig102}. Obviously,
$A_{dyna, sext}$ scales with $1/S$.
Fig. \ref{fig119} gives the similar comparison for an octupole located at 
$s=s_2$, and confirms that $A_{dyna, oct}$  scales with $1/\sqrt{S}$.
\item[8)] Now we change the tune of the machine a little bit (from $\nu_x=1.7$
to $\nu_x=1.565$), and the corresponding horizontal beta function distribution
and the third order tune diagram are shown in Figs. \ref{fig112} 
and \ref{fig113}. It is known that in this case $\beta_x(0)=5.1$ m.
A sextupole of $S=2$ is located a $s=s_1$ with 
$\beta_x(s_1)=12.42$ m, and its influence on the dynamic aperture is shown 
in Figs. \ref{fig114} and \ref{fig115}. From eq. \ref{eq:32} one finds
$A_{dyna,sext}=0.021$ m compared with the numerical
value of 0.02 m.
\item[9)] Finally, a 2D dynamic aperture is calculated numerically and analytically.
If a sextupole of $S=2$ is located at $s_2$ in 
the same lattice as in case (1) with
$\beta_x=15.18$ m and $\beta_y=4.26$ m, the 2D dynamic aperture is calculated by using BETA and eq. \ref{eq:12gg} as shown in Fig. \ref{fig300} and  
Fig. \ref{fig301}, respectively. The peak analytical
dynamic apertures in horizontal and vertical planes are 0.0163 m and 0.031 m,
compared with the numerical results of 0.017 m and 0.034 m, respectively.
\end{itemize}

\newpage
\begin{figure}[h]
\vskip 2.5 true cm
\vspace{6.0cm}
\includegraphics{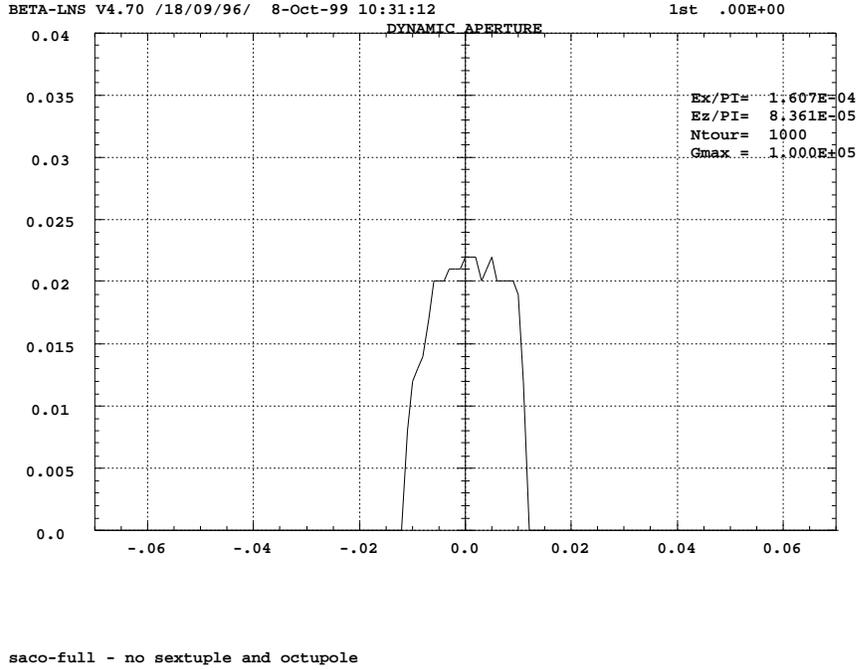}
\vskip 0. true cm
\caption{The dynamic aperture plot ($S(s_1)=2$, $O(s_2)=62$,
 $\beta_x(s_1)=13.6$ m, and  $\beta_x(s_2)=15.18$ m ).
\label{fig110}}
\end{figure}
\vspace*{40mm}
\begin{figure}[h]
\vskip 0.5 true cm
\vspace{6.0cm}
\includegraphics{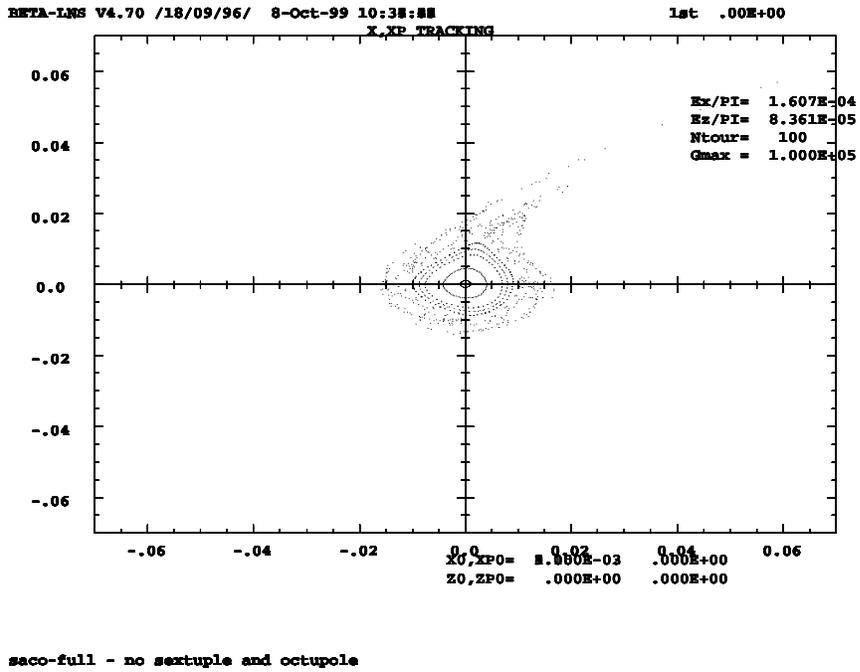}

\vskip 0. true cm
\caption{The horizontal phase space ($S(s_1)=2$, $O(s_2)=62$,
 $\beta_x(s_1)=13.6$ m, and  $\beta_x(s_2)=15.18$ m). 
\label{fig111}}
\end{figure}
\newpage
\begin{figure}[h]
\vskip 2.5 true cm
\vspace{6.5cm}
\includegraphics{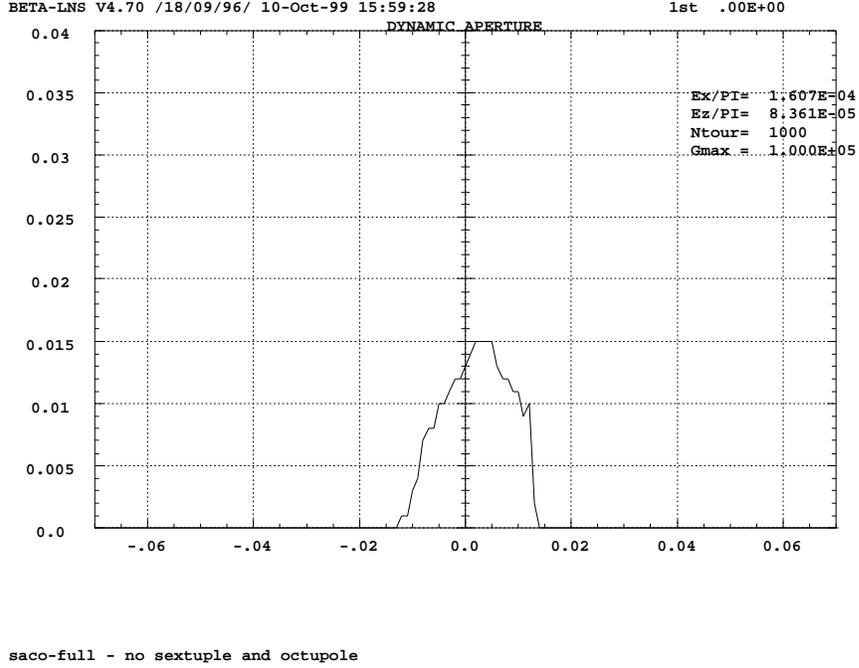}
\vskip 0. true cm
\caption{The dynamic aperture plot ($S(s_{1,2,3,4})=2$,
$\beta_x(s_1)=13.6$ m, $\beta_x(s_2)=15.18$ m,  
$\beta_x(s_3)=7.8$ m, and  $\beta_x(s_4)=6.8$ m).
\label{fig110a}}
\end{figure}
\vspace*{50mm}
\begin{figure}[h]
\vskip 0.5 true cm
\vspace{6.0cm}
\includegraphics{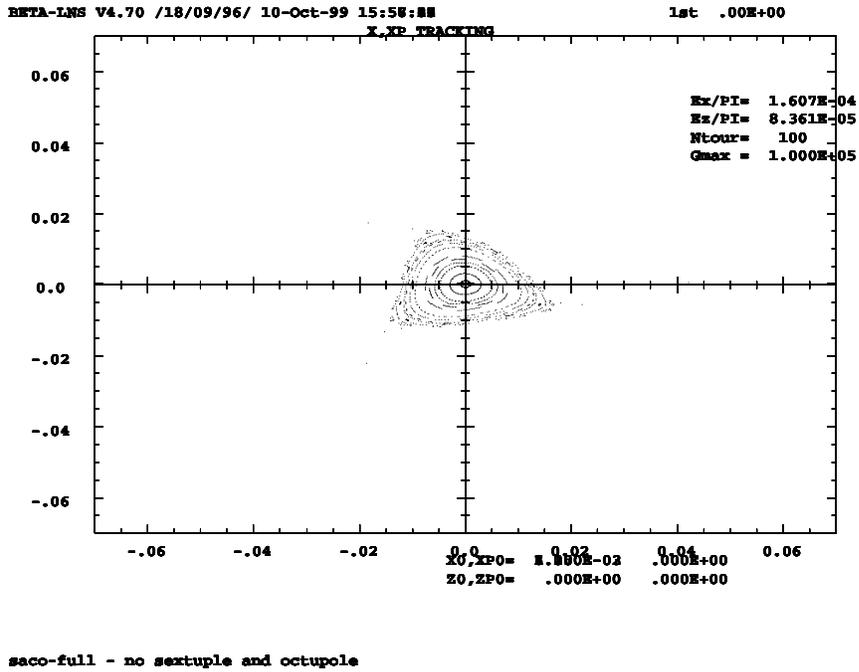}
\vskip 0. true cm
\caption{The horizontal phase space ($S(s_{1,2,3,4})=2$,
$\beta_x(s_1)=13.6$ m, $\beta_x(s_2)=15.18$ m,  
$\beta_x(s_3)=7.8$ m, and  $\beta_x(s_4)=6.8$ m). 
\label{fig111a}}
\end{figure}
\newpage
\begin{figure}[h]
\vskip 2.4 true cm
\vspace{7.0cm}
\includegraphics{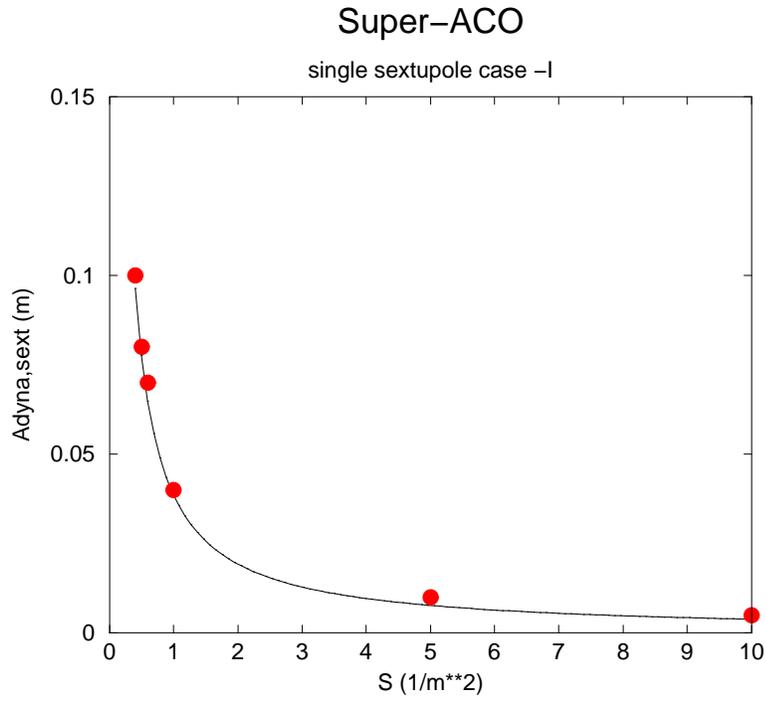}
\vskip -. true cm
\caption{The dynamic aperture of Super-ACO vs S ($S=b_2L/\rho$) at $s_1$.
\label{fig116}}
\end{figure}
\vspace*{40mm}
\begin{figure}[h]
\vskip 0. true cm
\vspace{7.0cm}
\includegraphics{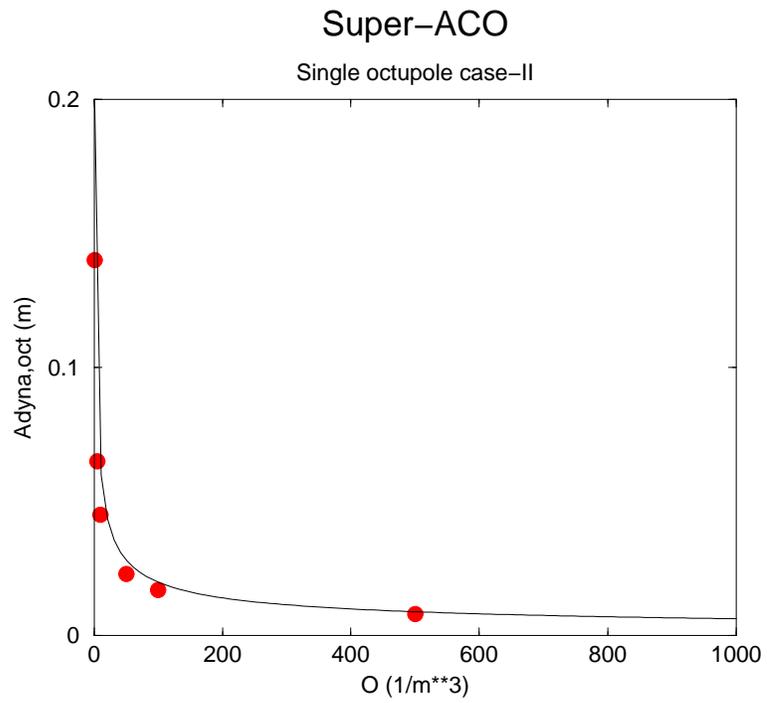}
\vskip -. true cm
\caption{The dynamic aperture of Super-ACO vs O ($O=b_3L/\rho$) at $s_2$.
\label{fig119}}
\end{figure}
\newpage
\begin{figure}[h]
\vskip 2.5 true cm
\vspace{7.0cm}
\includegraphics{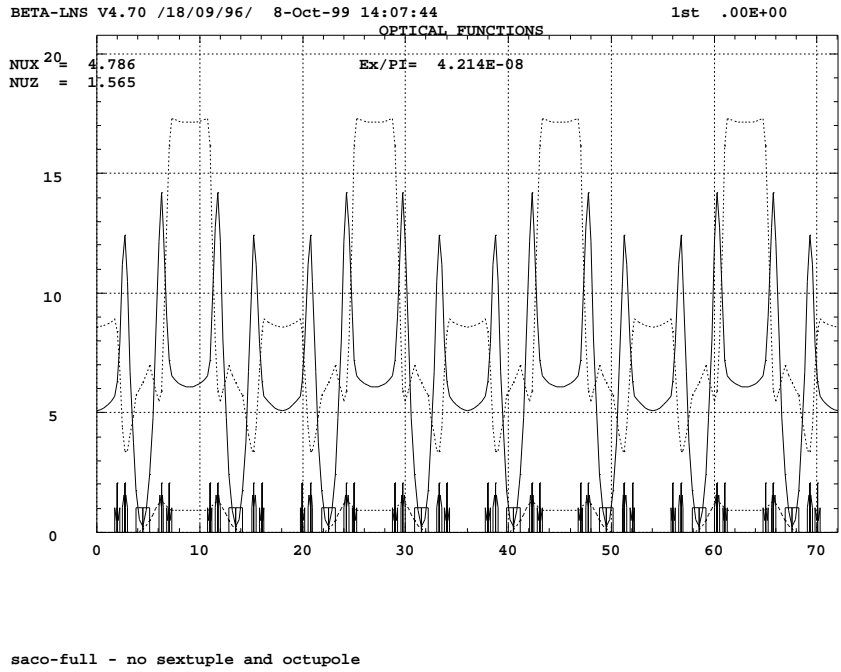}
\vskip 0. true cm
\caption{The horizontal beta function distribution of modified Super-ACO
($\beta_x(0)=5.1$ m).
\label{fig112}}
\end{figure}
\vspace*{40mm}
\begin{figure}[h]
\vskip 0. true cm
\vspace{6.0cm}
\includegraphics{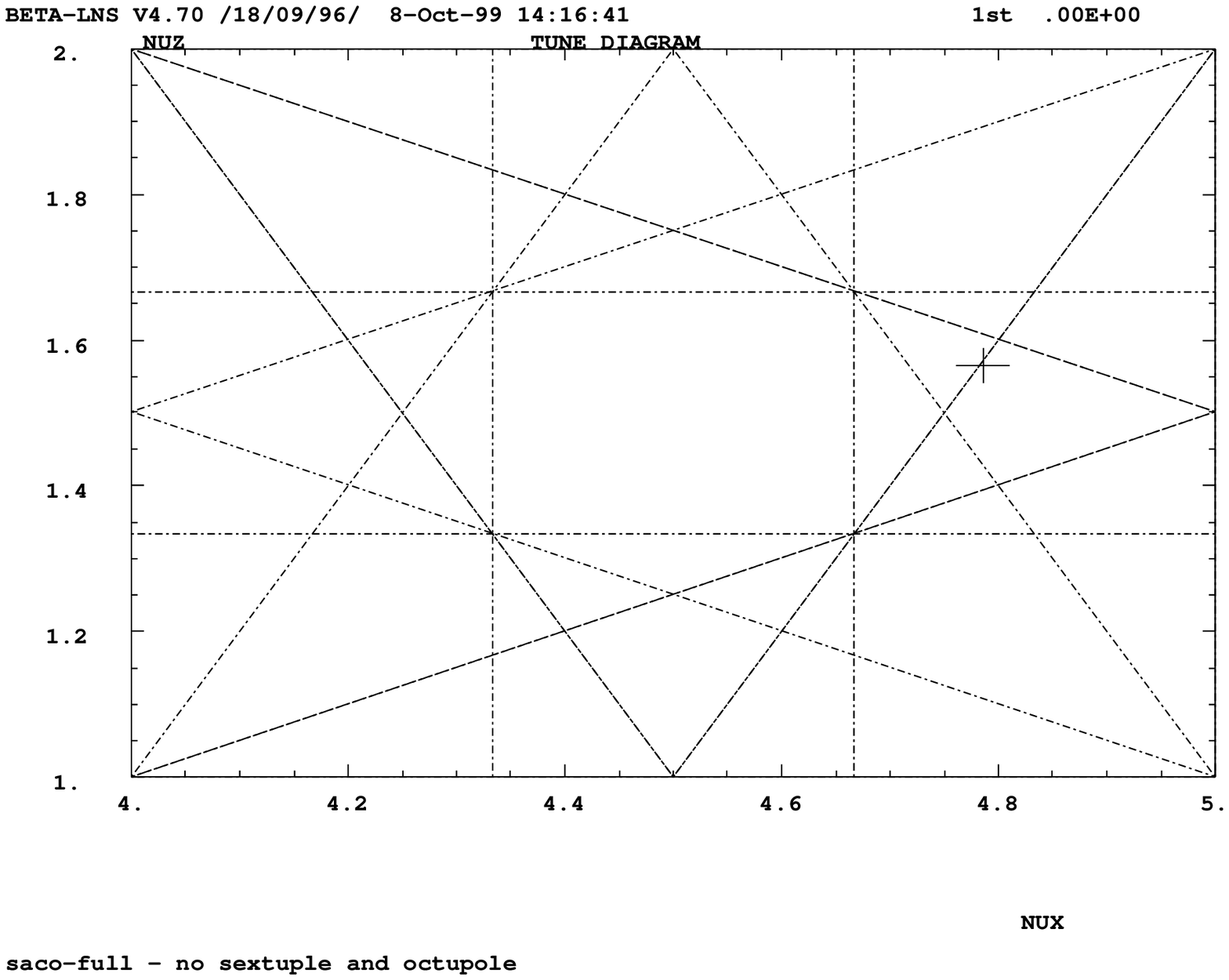}
\vskip 0. true cm
\caption{The tune diagram of the third order of modified 
Super-ACO, where the cross
indicates the working point of the machine.
\label{fig113}}
\end{figure}
\newpage
\begin{figure}[h]
\vskip 2.5 true cm
\vspace{7.0cm}
\includegraphics{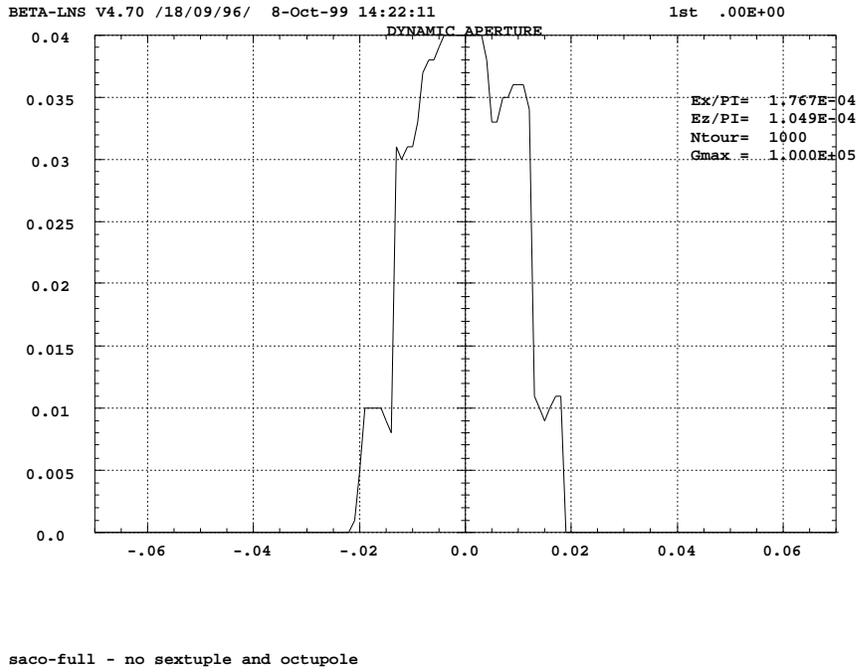}
\vskip 0. true cm
\caption{The dynamic aperture plot ($S(s_1)=2$ and $\beta_x(s_1)=12.42$ m).
\label{fig114}}
\end{figure}
\vspace*{50mm}
\begin{figure}[h]
\vskip 0. true cm
\vspace{6.0cm}
\includegraphics{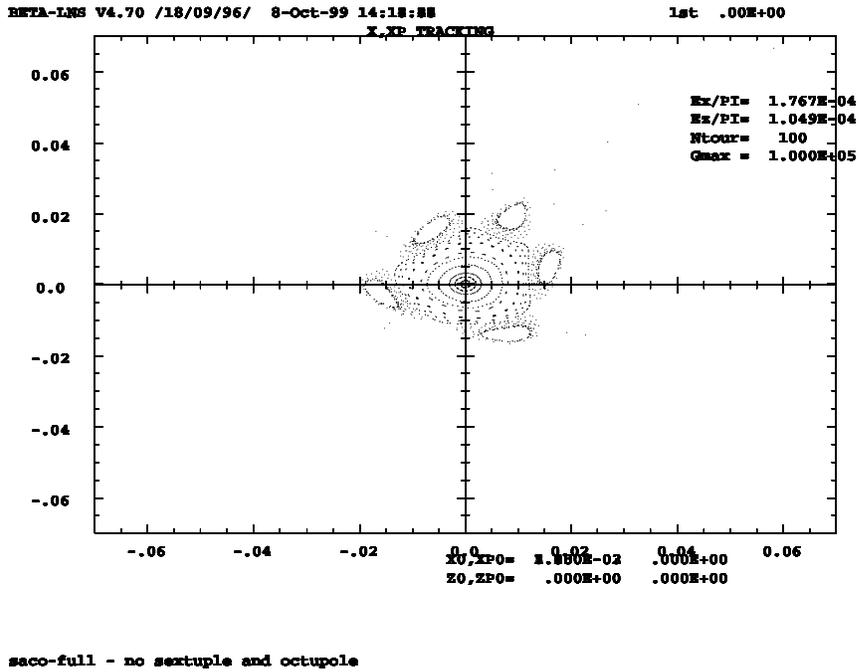}
\vskip 0. true cm
\caption{The horizontal phase space ($S(s_1)=2$ and
$\beta_x(s_1)=12.42$ m). 
\label{fig115}}
\end{figure}
\newpage
\begin{figure}[h]
\vskip 2.5 true cm
\vspace{6.0cm}
\includegraphics{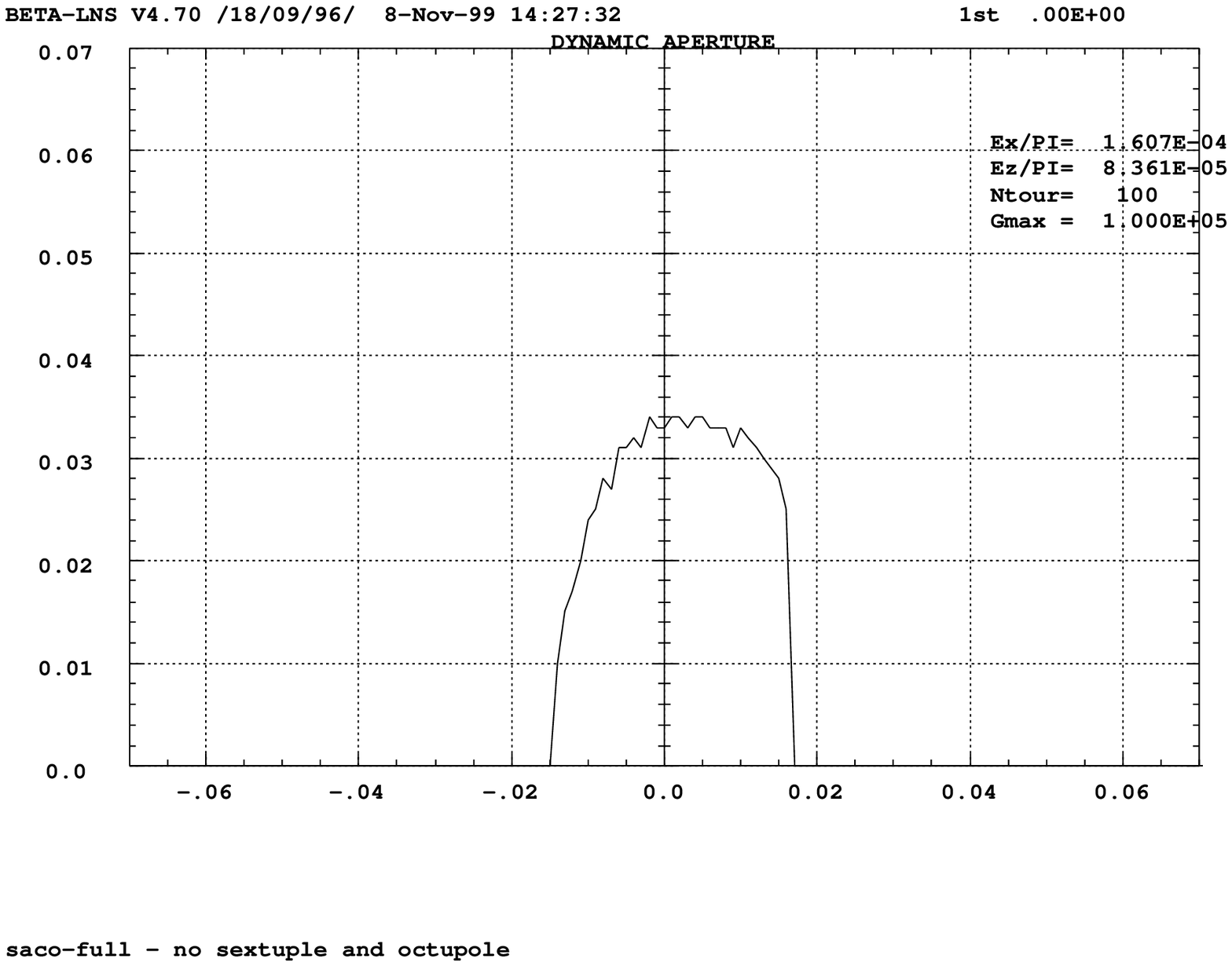}

\vskip 0. true cm
\caption{The 2D dynamic aperture of Super-ACO with S=2 located
at $s_2$ with $\beta_x(s_2)=15.18$ m and $\beta_y(s_2)=4.26$ m.
\label{fig300}}
\end{figure}
\vspace*{40mm}
\begin{figure}[h]
\vskip 0.5 true cm
\vspace{6.0cm}

\includegraphics{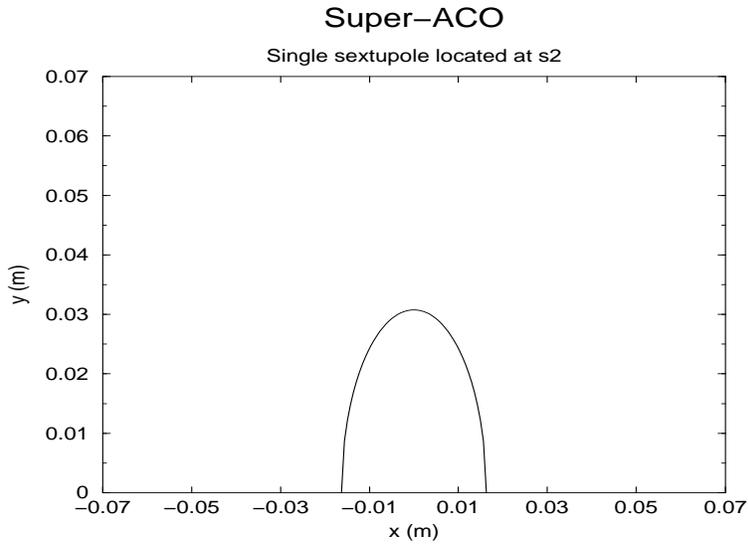}
\vskip -. true cm
\caption{The analytical estimation of the 2D dynamic aperture of Super-ACO
 with S=2 located at $s_2$ with $\beta_x(s_2)=15.18$ m and
 $\beta_y(s_2)=4.26$ m.
\label{fig301}}
\end{figure}

\newpage
To make the comparison much more clear we illustrate the machine parameters
in Table 1 and the comparison results in Table 2.

\par
The agreement between the analytical and the numerical 
simulation results is quite good. The 
fact that the analytically estimated 2D dynamic aperture agrees well with that
calculated by the numerical method shows that the
method we have used to
treat the two coupled nonlinear oscillators is 
a reasonable heuristic procedure.    

\begin{table}[h]
\begin{center}
\begin{tabular}{|l|l|l|}
\hline
Case&Multipole strength&beta function (m)\\
\hline
1&$S(s_1)=2$ (1/m$^2$)&$\beta_x(s_1)=13.6$ \\
\hline
2&$O(s_1)=10$ (1/m$^3$)&$\beta_x(s_1)=13.6$ \\
\hline
3&$D(s_1)=1000$ (1/m$^4$)&$\beta_x(s_1)=13.6$ \\
\hline
4&$S(s_1)=2$ (1/m$^2$), $O(s_1)=62$ (1/m$^3$)&$\beta_x(s_1)=13.6$ \\
\hline
5&$S(s_1)=2$ (1/m$^2$), $O(s_2)=62$ (1/m$^3$)&$\beta_x(s_1)=13.6$ , $\beta(s_2)=15.18$ \\
\hline
6&$S(s_{1,2,3,4})=2$ (1/m$^2$)&$\beta_x(s_{1,2,3,4})=$13.6, 15.18, 7.8, 6.8 \\
\hline
8&$S(s_1)=2$ (1/m$^2$)&$\beta_x(s_1)=12.42$ , $\beta_x(0)=5.1$ \\
\hline
9&$S(s_1)=2$ (1/m$^2$)&$\beta_x(s_2)=15.18$ \\
\hline
\end{tabular}
\end{center}
\caption{summary of parameters}
\label{tab:1}
\end{table}
\begin{table}[h]
\begin{center}
\begin{tabular}{|l|l|l|}
\hline
Case&$A_{dyna,analy.}$ (m)
&$A_{dyna,numer.}$ (m)\\
\hline
1&$0.0385$ &
$0.04$\\
\hline
2&0.055 &
$0.054$ \\
\hline
3&$0.022$ &
$0.024$ \\
\hline
4&$0.0145$ &
$0.016$ \\
\hline
5 
&$0.0138$ &
$0.0135$ \\
\hline
6&$0.012$ &
$0.0135$ \\
\hline
8&$0.021$&
$0.02$\\
\hline
9&
$A_x=0.0163$ , $A_{y}=0.031$ &
$A_{x}=0.017$, $A_{y}=0.034$ \\
\hline
\end{tabular}
\end{center}
\caption{summary of comparison results}
\label{tab:2}
\end{table}

\section{Beam-beam interaction limited dynamic apertures}
Since we are interested in the single particle motion under beam-beam forces,
the incoherent force should be taken into account \cite{Keil}:
\begin{equation}
F_{r,in}(r)=\pm {nq^2(1+\beta^2)\over 2 \pi \epsilon_0r}\left(
1-\exp\left(-{r^2\over 2 \sigma^2}\right)\right)
\label{eq:1aa}
\end{equation} 
where $n$ is the line particle number density, $\beta$ is the particle's
velocity in units of the speed of light, $q$ is the particle's electric charge,
$\epsilon_0$ is the permitivity in vacuum, $r$ is the transverse offset of a
particle with respect to the center of the counter-rotating colliding bunch,
$\sigma$
is the standard deviation of the Gaussian transverse charge density distribution, the positive and the negative signs correspond to the colliding bunches with the same or opposite charges, respectively.
Now we expand $F_{r,in}(r)$ into Taylor series:
\begin{equation}
F_{r,in}(r)=\pm {nq^2\over \pi \epsilon_0}\left(
{1\over 2\sigma^2}r-{1\over 8\sigma^4}r^3+{1\over 48\sigma^6}r^5
-{1\over 384\sigma^8}r^7
+\cdot \cdot \cdot \right)
\label{eq:2aa}
\end{equation}
where we take $\beta =1$. To start with, we consider the particle's
motion in horizontal plane ($y=0$) and consider only the delta function
nonlinear beam-beam forces
coming from one IP. The differential equation of motion can be expressed as: 
$${d^2x\over ds^2}+K_x(s)x=\pm {nq^2\over \pi \epsilon_0m_0c^2 \gamma}(
{1\over 2\sigma^2}x-{1\over 8\sigma^4}x^3+{1\over 48\sigma^6}x^5$$
\begin{equation}
-{1\over 384\sigma^8}x^7
+\cdot \cdot \cdot )L\sum_{k=-\infty}^{\infty}\delta(s-kL)
\label{eq:3aa}
\end{equation}
where $K_x(s)$ describes the linear focusing of the lattice in the horizontal plane, $m_0c^2$
is the rest energy of the particle, $\gamma$ is the normalized particle's
energy, and $L$ is the circumference of the circular collider.
The corresponding Hamiltonian is expressed as:
$$H={p_x^2\over 2}+{K_x(s)\over 2}x^2\mp {nq^2\over \pi \epsilon_0m_0c^2 \gamma}
({1\over 4\sigma^2}x^2-{1\over 32\sigma^4}x^4+{1\over 288\sigma^6}x^6$$
\begin{equation}
-{1\over 3072\sigma^8}x^8
+\cdot \cdot \cdot )L\sum_{k=-\infty}^{\infty}\delta(s-kL)
\label{eq:433}
\end{equation}
where $p_x=dx/ds$.
\par
To make use of the general dynamic aperture formulae shown in section 3,
one needs only to find the equivalence relations by comparing two Hamiltonians
expressed in eqs. \ref{eq:12} and \ref{eq:433}, respectively, and it is found 
that:
\begin{equation}
{b_3\over \rho}L={N_eq^2\over 8\pi \epsilon_0m_0c^2\gamma \sigma^4}
\label{eq:11aa}
\end{equation}
\begin{equation}
{b_5\over \rho}L={N_eq^2\over 48\pi \epsilon_0m_0c^2\gamma \sigma^6}
\label{eq:12aa}
\end{equation}
\begin{equation}
{b_7\over \rho}L={N_eq^2\over 384\pi \epsilon_0m_0c^2\gamma \sigma^8}
\label{eq:13aa}
\end{equation}
and so on, where we have replaced $nL$ by $N_e$ which is the particle
population inside a bunch. Till now one 
can calculate all kinds of dynamic apertures due to nonlinear beam-beam
forces.
For example, one can get the dynamic apertures due to the beam-beam
octupole nonlinear force:
$$
A_{dyna,8,x}={\sqrt{\beta_x(s)}\over \beta_x(s^*)}
\sqrt{\rho \over \vert b_3\vert L}$$
\begin{equation}
={\sqrt{\beta_x(s)}\over \beta_x(s^*)}\left({8\pi 
\epsilon_0m_0c^2\gamma \sigma^4\over N_eq^2}\right)^{1/2} 
\label{eq:14aa}
\end{equation}
and 
\begin{equation}
A_{dyna,8,y}=\sqrt{{\beta_x(s^*)\over \beta_y(s^*)}(A^2_{dyna,8,x}-x^2)}
\label{eq:15aa}
\end{equation}
where $s^*$ is the IP position. If we measure dynamic apertures by the
beam sizes, one gets:
\begin{equation}
{\cal R}_{x,8}={A_{dyna,8,x}\over \sigma_x(s)}
=\left({8\pi 
\epsilon_0m_0c^2\gamma \epsilon_x\over N_eq^2}\right)^{1/2} 
\label{eq:16aa}
\end{equation}
where $\epsilon_x$ is the bunch horizontal emittance.
When the higher order multipoles effects ($2m >8$) can be neglected 
eqs. \ref{eq:14aa} and \ref{eq:15aa} give very good approximations to the 2D
dynamic apertures limited by one beam-beam IP. If there are $N_{IP}$ interaction points in a ring the dynamic apertures described in eqs. \ref{eq:14aa} and
\ref{eq:15aa} will be reduced by a factor of $\sqrt{N_{IP}}$ (if these $N_{IP}$
interaction points can be regarded as independent). Given the dynamic
aperture of the ring without the beam-beam effect as $A_{x,y}$, 
the total dynamic aperture including the beam-beam effect can be 
estimated usually as:     
\begin{equation}
A_{total,x,y}={1\over \sqrt{{1\over A_{x,y}^2}+{1\over A_{bb,x,y}^2}}}
\label{eq:18aa}
\end{equation} 
\par
Taking PEP-II B-Factory design parameters for example \cite{Seeman} and
assuming that the beams are round at IP, 
for the high energy ring, $N_e=2.8\times 10^{10}$, $\gamma=1.76\times 10^4$, and $\epsilon_x=49$ nm,
one gets from eq. \ref{eq:16aa}, ${\cal R}_{x,8}=3.2$, and for the low energy ring, $N_e=6\times 10^{10}$, $\gamma=6.07\times 10^3$, and $\epsilon_x=49$ nm,
one finds ${\cal R}_{x,8}=2.7$. 
\par
\section{Conclusion}
We have derived the analytical formulae for the dynamic 
apertures in circular accelerators due to single sextupole, single octupole,
single decapole (single 2$m$ pole in general),
and the combination of many independent multipoles.
The analytical results have been systematically compared with the 
numerical ones and the agreement is quite satisfactory.   
These formulae are very useful both for the physical insight and in the
practical machine design and operation. One application of these formulae
is to estimate analytically the beam-beam interaction determined 
dynamic aperture in a circular collider.
\par
\section{ Acknowledgements}
The author of this paper thanks B. Mouton for his help in using BETA program
and also for his generating a flexible lattice based on the original
Super-ACO one. The fruitful discussions with A. Tkachenko are 
very much appreciated. 
\par


\begin{thebibliography}{11}
\bibitem{1} R. Ruth, ``Single particle dynamics in circular accelerators'',
AIP conference proceedings, No. 153, p. 150.
\bibitem{2} B.W. Montague, ``Basic Hamiltonian mechanics'', CERN 95-06, Vol. I,
p. 1.
\bibitem{3} E.J.N. Wilson, ``Nonlinear resonances'', CERN 95-06, Vol. I, p. 15.
\bibitem{4} J.S. Bell, ``Hamiltonian mechanics'', CERN 87-03, p. 5.
\bibitem{5} A.A. Kolomesky and A.N. Lebedev, 
``Theory of cyclic accelerators'', North-Holland Publishing Comp. (1966).
\bibitem{6} T. Suzuki, ``Hamiltonian formulation for synchrotron
oscillations and Sacherer's integral equation'', {\it Particle Accelerators},
Vol. 12 (1982), p. 237.
\bibitem{7} H. Wiedemann, ``Particle Accelerator Physics II'',
 Springer-Verlag, 1993. 
\bibitem{9} R.Z. Sagdeev, D.A. Usikov, and G.M. Zaslavsky, ``Nonlinear Physics,
from the pendulum to turbulence and chaos'', Harwood Academic Publishers, 1988.
\bibitem{10} B.V. Chirikov, ``A universal instability of many-dimensional
oscillator systems'', Physics Reports, Vol. 52, No. 5 (1979), p. 263.
\bibitem{11} J.M. Greene, J. Math. Phys. 20 (1979), p. 1183.
\bibitem{11a} A.J. Lichtenberg and M.A. Lieberman, ``Regular and stochastic
motion'', Springer-Verlag (1983), p. 46.
\bibitem{12} L. Farvaque, J.L. Laclare, and A. Ropert, ``BETA users' guide'',
ESRF-SR/LAL-88-08.
\bibitem{Keil} E. Keil, ``Beam-beam dynamics'', CERN 95-06, p. 539.
\bibitem{Seeman} J.T. Seeman, ``Commissioning results of the KEKB and PEP-II
B-Factories'', Proceedings of PAC'99 (1999), p. 1.
\end{thebibliography}
\end{document}